\documentstyle[12pt]{article}
\begin{document}

\def\theequation{\arabic{section}.\arabic{equation}}
\def\Section#1{\setcounter{equation}{0}\section{#1}}
\def\ar{\rightarrow}
\def\deh{{\pa}_{0}^{\!\!\!\!\leftrightarrow}\,}
\def\disc{\mbox{\,disc}\,}
\def\intf{\int d^{4}x\,}
\def\intgr{\int_{4}^{\infty}}
\def\ints{\int_{4}^{\infty} ds'}
\def\lar{\longrightarrow}
\def\pa{\partial}
\def\pal{p_{\al}}
\def\pbe{p_{\be}}
\def\pga{p_{\ga}}
\def\pde{p_{\de}}
\def\qba{\overline{q}}
\def\roo{(\frac{s-4}{4})}
\def\suz{\mbox{SU(2)}}
\def\sud{\mbox{SU(3)}}
\def\tp{\tilde p}
\def\tr{\mbox{tr}\,}
\def\Tr{\mbox{Tr}\,}
\def\ue{\mbox{U(1)}}
\def\uka{\underline{k}}
\def\upe{\underline{p}}
\def\al{\alpha}
\def\be{\beta}
\def\ga{\gamma}
\def\de{\delta}
\def\ka{\kappa\,}
\def\la{\lambda}
\def\ep{\varepsilon}
\def\om{\omega}
\def\Ph{{\it\Phi}}
\def\psiba{\overline{\psi}}
\def\si{\sigma}
\def\th{\theta}
\def\va{\varphi}

\def\ar{\rightarrow}\
\def\disc{\mbox{\,disc}\,}
\def\intf{\int d^{4}x\,}
\def\intgr{\int_{4}^{\infty}}
\def\ints{\int_{4}^{\infty} ds'}
\def\lar{\longrightarrow}
\def\pa{\partial}
\def\pal{p_{\al}}
\def\pbe{p_{\be}}
\def\pga{p_{\ga}}
\def\pde{p_{\de}}
\def\qba{\overline{q}}
\def\roo{(\frac{s-4}{4})}
\def\suz{\mbox{SU(2)}}
\def\sud{\mbox{SU(3)}}
\def\tp{\tilde p}
\def\tr{\mbox{tr}\,}
\def\Tr{\mbox{Tr}\,}
\def\ue{\mbox{U(1)}}
\def\uka{\underline{k}}
\def\upe{\underline{p}}
\def\al{\alpha}
\def\be{\beta}
\def\ga{\gamma}
\def\de{\delta}
\def\ka{\kappa\,}
\def\la{\lambda}
\def\ep{\varepsilon}
\def\om{\omega}
\def\Ph{{\it\Phi}}
\def\psiba{\overline{\psi}}
\def\si{\sigma}
\def\th{\theta}
\def\va{\varphi}

\def\beq{\begin{equation}}
\def\eeq{\end{equation}}
\def\bed{\begin{displaymath}}
\def\eed{\end{displaymath}}
\def\beqq{\begin{eqnarray}}
\def\eeqq{\end{eqnarray}}
\def\bedd{\begin{eqnarray*}}
\def\eedd{\end{eqnarray*}}
\def\Ph{{\it\Phi}}


\begin{titlepage}
\begin{flushright}
ZU-TH 41/94\\
IPNO/TH 94-93\\
STP-DIAS-95-23
\end{flushright}
\vfill
\centerline{\Large\bf Final State Interactions and Khuri-Treiman }
\centerline{\Large\bf Equations in $\eta\to 3\pi$ decays
$\footnote{Work supported in part by
    Schweizerischer Nationalfonds.}$}
\vfill
\centerline{\large J. Kambor$^a$, C. Wiesendanger $^{b,c}$, D. Wyler$^b$}
\vspace*{0.5cm}
 \vspace*{0.5cm}
\centerline{$^a$
 Division de Physique Th\'eorique
$\footnote{Unit\'e de Recherche des Universit\'es Paris 11 et Paris 6,
associ\'ee au CNRS.}$, Institut de Physique Nucl\'eaire,}
\centerline{F-91406 Orsay Cedex, France}
\centerline{$^b$ Theoretische Physik, Universit\"at Z\"urich, CH-8057
  Z\"urich, Switzerland}
\centerline{$^c$ Dublin Institute for Advanced Studies, School of
    Theoretical Physics,}
\centerline{10 Burlington Road, Dublin 4, Ireland}
\vfill
\begin{abstract}
Using extended Khuri-Treiman equations, we evaluate
the final state interactions
due to two-pion rescatterings to the decays
$\eta\to \pi^0 \pi^+ \pi^-$ and $\eta\to \pi^0 \pi^0 \pi^0$.
As subtraction to the dispersion relation
we take the one-loop chiral perturbation theory
result of Gasser and Leutwyler. The calculated corrections are moderate
and amount to about $14\%$ in the amplitude at the center of the decay region.
A careful analysis of the errors inherent to our approach is given.
As a consequence, the experimental rate of the decay can only be reproduced
if the double quark mass ratio $Q^{-2}\equiv\frac{m_{d}-m_{u}}
{m_{s}-{\hat m}}
\cdot\frac{m_{d}+m_{u}}{m_{s}+{\hat m}}$ is increased
from the usual value of $1/(24.1)^2$ to $1/(22.4 \pm 0.9)^2$.
We have also calculated the ratio of the rates of the two decays and
various Dalitz Plot parameters. In particular, the linear slope $a$ in
the charged decay is different from the one-loop value and
agrees better with experiment.
\newline
\hspace{4cm}
\newline
\end{abstract}

\end{titlepage}
\newpage

\Section{Introduction}
\paragraph{}
Chiral perturbation theory (ChPT) \cite{CH1,CH2} offers a
consistent
description of low energy QCD, in particular of the
strong and
weak mesonic interactions. Whereas most phenomena are
quite well
accounted for,
the predictions for the decays $\eta\to 3\pi$ remain
much below the experimental results. The lowest order
amplitude is fully determined by chiral symmetry \cite{OW} to be
\beqq \label{amp} F(s_{a},s_{b},s_{c})&=&-\frac{B_{0}
(m_{d}-m_{u})}{3{\sqrt3} F_{0}^2}\cdot f^{(2)}(s_{a}) \\
f^{(2)}(s_{a})&\equiv&T(s_a)=1
+\frac{3(s_{a}-s_{0})}{m_{\eta}^{2}-m_{\pi}^{2}} \nonumber
\eeqq
where $m_u$ and $m_d$ are the up and down quark masses,
$F_0$ and $B_0$ QCD parameters and $s_{a},s_{b},s_{c}$,
$s_0$ kinematical variables to be defined later. While
the form of eq. (\ref{amp}) is valid for the decay into
$\pi^0 \pi^+ \pi^-$, the expression for three neutral
pions is obtained by setting $f^{(2)}(s_{a})=1$.
Furthermore, in lowest order in the chiral expansion
one has
\beqq \label{massrel}B_{0}(m_{d}-m_{u})&=&
(m_{K^{0}}^{2}-m_{K^{+}}^{2})-(m_{\pi^{0}}^{2}-m_
{\pi^{+}}^{2}) \\
&\equiv&m_{1}^{2}. \nonumber \eeqq
These equations yield a width of $66$ eV for the
decay $\eta \to \pi^0 \pi^+ \pi^-$ which is far
below the experimental value of $281 \pm28$ eV \cite{DAT}.
It is an unusual situation, since most predictions
of current algebra fall within $20$ to $30$\% of the
experimental results.

However, also the next-to-leading calculation of
Gasser and Leutwyler \cite{GL85c}
failed to reproduce the experimental value, despite
a dramatic improvement. These authors obtain a width
of $167 \pm50 $eV for the $\eta \to \pi^0 \pi^+ \pi^-$ decay
(We have adopted the 'new' value of $f_\pi$
\cite{fpi} which increases
the width by about $7$ eV).
Thus, one may conclude that higher order corrections are
very large, possibly preventing a satisfactory representation
of this decay by ChPT.

Since the decay rate is proportional to $(m_u - m_d)^2$
it is particularly sensitive to the value of the quark masses.
The quark mass ratios are known with some precision \cite{MASS}
from a variety of low energy investigations. Since they
are fundamental parameters of the basic theory, they
should be determined as accurately as possible. Of
special interest is the up quark mass because $m_u = 0$
is a very appealing solution to the strong CP problem  \cite{PQ} .
 From eq. (\ref{amp}) we see that decreasing $m_u$ indeed
increases the rate as
required by experiment. On the other hand, corrections
to the second factor will also have this effect. Thus,
only a careful investigation of higher order effects
will enable one to draw a conclusion on the quark masses.

Let us formulate this more precisely. The amplitude
for the decay $\eta \to \pi^0 \pi^+ \pi^-$ can
be written as  \cite{GL85c}
\beq \label{amptot} F(s_{a},s_{b},s_{c})=
-\frac{B_{0}(m_{d}-m_{u})}{3{\sqrt3} F_{0}^2} \cdot
f(s_{a},s_{b},s_{c},{\it\Lambda},m_{u},..)+\mbox{e.m.} \eeq
where the quark masses are the renormalization group
invariant masses,
${\it\Lambda}$ a QCD scale, $s_{a},s_{b},s_{c}$ the
Mandelstam variables for the decay. Sutherland's
theorem  \cite{SUT}  implies that the electromagnetic
contribution is of order $p^2$, where $p^2$
stands for any invariant product of momenta.
Moreover, the leading (in momentum) electromagnetic
contribution of order $\frac{e^2 p^2}{{\it\Lambda}^2}$
is further suppressed by a factor $\frac
{m_\pi^2}{m_K^2}$ if the amplitude is
assumed to be linear in $s_a$ \cite{BS68,DDE73}.
Recently, these terms
were reanalyzed in an effective Lagrangian
framework \cite{BKW} and the
expected smallness of the
electromagnetic corrections (Sutherland's theorem)
was confirmed.
We will therefore neglect them throughout.

Returning then to the QCD contribution in
eq. (\ref{amptot}),
the second factor, the function $f$,
is expanded in chiral perturbation theory in powers
of momentum and mass
\beq \label{chiramp} f=f^{(2)}+f^{(4)}+.. \eeq
Here, $f^{(2)}$ has already been defined in eq. (\ref{amp})
and $f^{(4)}$, etc.
are of higher orders. $f^{(4)}$ was given explicitely
by Gasser and Leutwyler \cite{GL85c} and consists of a variety
of loops and counterterms. The corresponding
coupling constants also occur in other
observables
which are also
calculated using the chiral Lagrangian, for
instance the
decay constants or the masses of the mesons. To
order $p^4$,
it is therefore possible to express some of these
corrections through
physical quantities. In particular, the quark
mass contribution to the kaon mass difference can
be written as
\cite{GL85a}
\beqq \label{meson}
 {\it\Delta}K_{QCD}\equiv(m_{K^{0}}^{2}-m_{K^{+}}^{2})_{QCD}
               = Q^{-2}\frac{m_K^2}{m_\pi^2}(m_K^2-m_\pi^2)
\eeqq
and can be obtained from the
measured masses once the
electromagnetic
mass differences are subtracted.
Here,
\beq \label{massdiff}Q^{-2}\equiv\frac{m_{d}-m_{u}}
{m_{s}-{\hat m}}
\cdot\frac{m_{d}+m_{u}}{m_{s}+{\hat m}} \nonumber \eeq
with ${\hat m}\equiv\frac{1}{2}(m_{d}+m_{u})$.

Then, one can rewrite the amplitude as
\beqq \label{truamp} F(s_{a},s_{b},s_{c})&=&-Q^{-2}\frac
{m_K^2}{m_\pi^2}(m_K^2-m_\pi^2)
\frac{1}{3{\sqrt 3}F_{\pi}^2} \\
& & \,\,\cdot f^{(2)}(s_{a})(1+\delta_{DEC}).
\nonumber \eeqq
$\delta_{DEC}$ contains the remaining
part of the $O(p^4)$ terms as well as corrections
from higher orders in the chiral expansion.
Terms of order $(m_{d}-m_{u})^2$ are tiny and
are therefore neglected throughout. As anticipated,
the rate is indeed proportional to
the quark mass difference squared.
\footnote{In passing, we note that in the generalized framework
of chiral perturbation theory \cite{GCHPT}, matters
are more complicated and a careful reanalysis of
all relations entering eq. (\ref{truamp}) is
required.}

The decay  $\eta\to 3\pi$
fixes the quantity $Q^2$ rather than the ratio
$\frac{m_u}{m_d}$. This is in agreement with the
observation by Kaplan and Manohar \cite{KM} that the
chiral Lagrangian possesses a curious reparametrization
invariance under which masses and some of the other
parameters are changed. It implies that matrix
elements derived from Green functions which do not
involve scalar or pseudoscalar densities are
invariant \cite{MASS2} with respect to that transformation. And
indeed, the combination $Q^2$ in eq. (\ref{massdiff})
is invariant, while for instance $\frac{m_u}{m_d}$ is not.
Unfortunately, there is no direct
experimental access to this ratio, the relevant Green
functions must be obtained from semiphenomenological
analyses \cite{MASS}.

There exists an independent determination of $Q^2$
by means of the following equation
\beq \label{newneun}
m_1^2 = {\it\Delta}K_{QCD} + (\Delta K-\Delta \pi)_{em}.
\eeq
Provided that the electromagnetic contribution
is known, eqs. (\ref{meson})
and (\ref{newneun})
can be used to check the consistency of the
different calculations or to predict the
rate for the decay $\eta\to 3\pi$.

The calculation of the electromagnetic
mass differences is notoriously difficult. Dashen's
theorem \cite{DT} implies that the electromagnetic
mass difference
of the kaons is equal to the mass difference
of the pions, \footnote{the QCD contribution to
the pion mass difference is proportional to
$(m_{d}-m_{u})^2$ and negligible} up to terms
of order $e^2p^2$. If these are
neglected,
one has  ${\it\Delta}K_{QCD}=m_{1}^{2}$
and one can write

\beq \label{betamp} F(s_{a},s_{b},s_{c})=
-\frac{m_{1}^{2}}{3{\sqrt 3}F_{\pi}^2}
\cdot f^{(2)}(s_{a})(1+\delta_{DEC})
\eeq
and $\delta_{DEC}$ determines the rate.
In their calculation of the decay rate for $\eta\to 3\pi$
Gasser and Leutwyler \cite{GL85c} obtained a
correction $\delta_{DEC}$
of about $50$\%. Together with eq. (\ref{betamp}), the value
of $167$ eV for the width resulted, still far below the
experimental number.

However, it was found recently \cite{DHW,RU}
that the corrections to
Dashen's theorem could be
substantial. One should therefore start from
eq. (\ref{truamp}),
avoiding unnecessary uncertainties arising from
corrections to Dashen's theorem. The normalization
of the  $\eta\to 3\pi$ amplitude is then determined
by $Q^2$ which follows in principle
from eq. (\ref{meson}). As pointed out in
\cite{DHWL}, the positive corrections to Dashen's theorem found in \cite{DHW}
increase ${\it\Delta}K$ such that
together with the value of  $\delta_{DEC}$ obtained
by Gasser and Leutwyler the experimental rate
for $\eta\to 3\pi$ would be reproduced.

Thus, the rate
of the $\eta$ decay can be increased either by determining
further corrections in $\delta_{DEC}$ (beyond
those calculated by Gasser and Leutwyler) or by increasing
the value of
${\it\Delta}K_{QCD}$. In view of eq. (\ref{massdiff}) this latter
term amounts to an increase in $Q^2$ and thus to a smaller
up quark mass.
But before any conclusion on the quark masses
can be drawn, the decay corrections must be understood better.

It has been argued \cite{Pich} that $\eta \eta'$ mixing may increase
the theoretical rate to the experimental value. However, it was recently
shown \cite{Heiri} that the effects of the $\eta'$ are in fact included in the
standard treatment and that the enhancement is due to an incomplete
treatment of the resonances as an explanation for the constants of the
chiral Lagrangian. We therefore omit this issue in the following.

The goal of this paper is to calculate one class of
corrections in $\delta_{DEC}$, the so called unitary
corrections which ensure that the decay
amplitude satisfies unitarity. These corrections are
the sum of certain diagrams, namely those describing
the rescattering of the final state particles to all
orders. It is believed that they dominate the
corrections, in particular since the final state interactions
of two pions in the $I=0$ $s$-wave channel is strong and
attractive. In fact, in the one-loop calculation
of Gasser and Leutwyler they yield the largest
 new contributions. They account for roughly
three quarters of the corrections, pion loops
contributing about $85$\% and kaons and etas
about $15$\% thereof, while other terms provide
the remaining quarter (of course, the division
of the corrections depends on the renormalization
scale chosen; however, physical arguments such as
resonance saturation of the counterterms indicate
that the
scale used for the above numbers is reasonable).

A method to evaluate the unitary corrections has been
given long ago by Khuri and Treiman \cite{KT60}.
Keeping only
two particle rescatterings
of pions (and neglecting $p$-waves) they derived
a set of dispersion relations, the Khuri-Treiman equation.
This paper was subsequently analyzed by several
authors \cite{AAG,GB,BK,KAK,BR,APP}.
In particular, Kacser \cite{KAK}
derived the correct prescription for the  analytical
continuation of the partial wave projections of the
decay amplitude. Bronzan \cite{BR} and later
also Neveu and Scherk \cite{NS70}
omitting also the $I=2$ rescatterings, solved
the Khuri-Treiman equation approximately.
Using the properties of the
Omn$\grave{\mbox{e}}$s function $D(s)$ entering
their approximate solution they argued that
the remaining terms are small.

Some time ago, Roiesnel and Truong \cite{RT81} reconsidered
the problem, using similar, although not
equivalent methods to those of Neveu and Scherk to calculate the
unitary corrections. These authors found a large enhancement
of the amplitude, enough to account for the observed rate.
The numerical results of this work were criticized in Ref. \cite{GL85c}
on physical grounds and  we confirm their findings.
We believe therefore that their result
is a overestimate of the unitary corrections.

In this work we will use (and generalize by the inclusion of $p$-waves
(see also \cite{Anis})) the
method of Khuri and Treiman to
determine numerically the unitary corrections.
It is organized as follows.

In section 2 we review the general structure of the decay amplitude
for  $\eta\ar3\pi$ and
give the necessary definitions.In section 3 we discuss the Khuri-Treiman
equations and the inclusion of the $I=1$ rescatterings. In section 4
we discuss the subtraction procedure, and in section 5 we
build up an iterative scheme for the numerical solution of the set of
coupled equations for the projected amplitudes $U,V,W$.
In section 6 we discuss our numerical results and the quality of our
iteration, and then in section 7 we draw the conclusions. Some appendices
contain necessary technical details.

\Section{Isospin decomposition}
In this section we discuss the general form of the amplitude for the decay
$\eta \ar 3\pi$. It serves to fix the notation, to give the isospin
decomposition
of the amplitude and to discuss its symmetry properties.

We denote the three pions in the decay $\eta\ar 3\pi$ by the letters a,b,c,
their corresponding four-momenta by $k_a$, $k_b$, $k_c$ and label
their isospins
 by $\alpha$, $\beta$, $\gamma$. The invariant masses of the pions are
then $k_a^2=k_b^2=k_c^2=m_\pi^2$ disregarding all mass differences at this
stage. The four momentum of the isospin singlet $\eta$ is denoted by $k_r$.

In the standard model of the strong, electromagnetic and weak interactions,
the decay $\eta\ar 3\pi$ proceeds either through the isospin breaking piece
of the QCD hamiltonian or through an operator of electromagnetic origin.
For both contributions, the interaction
Hamiltonian $H(0)$ is $\Delta I=1$. It is then
convenient to treat the $\eta$ as a spurious isospin triplet and to assume
isospin conservation in the decay. The corresponding isospin state is labeled
by $\rho$.

The amplitude for a general isospin assignment of the pions is
\beqq
M&=& < \alpha,k_a;\beta,k_b;\gamma,k_c|H(0)|k_r> \nonumber \\
&\equiv&i (2\pi)^4 \delta(k_a+k_b+k_c-k_r) M_{\alpha\beta\gamma;\rho}
(s_a,s_b,s_c).
\eeqq
The invariant amplitude has the decomposition
\beqq \label{2.2}
M_{\alpha\beta\gamma;\rho}(s_a,s_b,s_c)
&=& F(s_a,s_b,s_c)\delta_{\beta\gamma}\delta_{\alpha\rho} \nonumber\\
&&+F(s_b,s_c,s_a)\delta_{\alpha\gamma}\delta_{\beta\rho} \nonumber\\
&&+F(s_c,s_a,s_b)\delta_{\alpha\beta}\delta_{\gamma\rho} ,
\label{ampM}
\eeqq
where the isoscalar amplitude $F$ is a function of the Mandelstam variables
\beq
s_a=(k_r-k_a)^2,\quad s_b=(k_r-k_b)^2,\quad s_c=(k_r-k_c)^2.
\eeq
Of course only two of these variables are independent as they are related by
the on-shell condition
\beq
s_a+s_b+s_c=3 m_\pi^2+m_\eta^2\equiv 3 s_0.
\eeq
Bose statistics of the three pion system implies the symmetry
\beq
F(s_a,s_b,s_c)=F(s_a,s_c,s_b)
\eeq
such that the amplitude $M_{\alpha\beta\gamma;\rho}(s_a,s_b,s_c)$ remains
invariant under the exchanges of
\beqq
a&\rightleftharpoons& a \qquad b\rightleftharpoons c, \nonumber\\
b&\rightleftharpoons& b \qquad c\rightleftharpoons a, \nonumber\\
c&\rightleftharpoons& c \qquad a\rightleftharpoons b .
\eeqq
The amplitude for the decay into charged pions is then found to be
\beq
<\pi^+,\pi^-,\pi^0 |H(0)|\eta> =
i (2\pi)^4 \delta(k_a+k_b+k_c-k_r) F(s_a,s_b,s_c),
\eeq
whereas the amplitude for the decay into neutral pions is expressed as
\beqq
<\pi^0,\pi^0,\pi^0 |H(0)|\eta> &=&
i (2\pi)^4 \delta(k_a+k_b+k_c-k_r) \\
&&\left[F(s_a,s_b,s_c)+F(s_b,s_c,s_a)+F(s_c,s_a,s_b)\right]. \nonumber
\eeqq

\Section{Isospin sum and scalar dispersion relation}

We begin with the dispersion representation
for the decay amplitude eq. (\ref{2.2}) derived by Khuri and
Treiman \cite{KT60}
\beqq \label{3.1}M_{\al\be\ga;\rho}(s_{a},s_{b},s_{c})
&=&\frac{1}{\pi}\ints_{a}\frac{\disc _{|_{\mbox{a}}}
M_{\al\be\ga;\rho}(s'_{a},s_{b},s_{c})}{s'_{a}-s_{a}-i\ep} \nonumber \\
&+&\frac{1}{\pi}\ints_{b}\frac{\disc_{|_{\mbox{b}}}
M_{\be\ga\al;\rho}(s_{a},s'_{b},s_{c})}{s'_{b}-s_{b}-i\ep} \\
&+&\frac{1}{\pi}\ints_{c}\frac{\disc_{|_{\mbox{c}}}
M_{\ga\al\be;\rho}(s_{a},s_{b},s'_{c})}{s'_{c}-s_{c}-i\ep}.
\nonumber \eeqq
Eqn.(\ref{3.1}) expresses the connection between
causality and analyticity for the considered amplitude.
Unitarity gives an expression
for the discontinuity of $M_{\al\be\ga;\rho}$ in
$s_a$ for fixed $s_c$ and analogous ones for the
other two discontinuities in the form of a sum over
intermediate states involved in all the
rescattering processes. The intermediate state with
lowest mass contributing in this sum is the two-pion
state leading to a rescattering of two outgoing pions
in the decay.
As any other intermediate state such as a $\bar K K$ or a
$4\pi$ state will contribute to the discontinuity at a
much larger threshold it is a reasonable approximation to
drop them \cite{KT60}. Hence, we find
\beqq \label{3.2}& &2i\disc_{|_{\mbox{a}}}
M_{\al\be\ga;\rho}(s_{a},s_{b},s_{c})=
\frac{i}{2!}\sum_{\de,\ep}\int d\tp_{d}d\tp_{e} \\
& &\quad\quad(2\pi)^{4}\de(k_{r}-k_{a}-p_{d}-p_{e})
N^{*}_{\de\ep;\be\ga}(s_{a},s'_{d},s'_{e})\cdot
M_{\al\de\ep;\rho}(s_{a},s_{d},s_{e}) \nonumber \eeqq
where $p_d$ and $p_e$ are the four-momenta of the
intermediate pions $d,e$.
In eq. (\ref{3.2}) we have introduced the invariant pion-pion
scattering matrix element
\beqq  \label{3.3}
& & <\al,k_{a};\be,k_{b}|T|\ep,p_{e};\eta,p_{f}>
\\ & &\quad \equiv
i(2\pi)^{4}\de(k_{a}+k_{b}-p_{e}-p_{f})N_{\al\be;\ep\eta}
(s_{a},s'_{d},s'_{e})
\nonumber \eeqq
and the phase space measure for the intermediate pions
\beq d\tp\equiv \frac{d^{3}p}{(2\pi)^{3}2E},\quad\quad E=\sqrt{p^{2}+1} .
\eeq
The variables $s'_{d}$ and $s'_{e}$ are
defined by
\beqq
s'_{d}&=&(k_{b}-p_{d})^{2} \\
s'_{e}&=&(k_{b}-p_{e})^{2}  \nonumber \eeqq
Note also that all momenta are taken in units of $m_{\pi}$.
In the Khuri-Treiman approximation, the discontinuity
of the decay amplitude becomes thus just a product of
the pion scattering amplitude with the decay amplitude itself.

To perform the phase space integrals in eq. (\ref{3.2}), one
usually chooses the c. m. frame of the rescattered pions defined by
$\uka_{b}+\uka_{c}=0$. Fixing the $\th$-angle in $d\tp_{d}$
to be the angle $\th_{da}$ between $\uka_{a}$ and $\upe_{d}$
the discontinuity becomes
\beqq \label{3.5}& & \disc_{|_{\mbox{a}}} M_{\al\be\ga;\rho}
=\frac{1}{32\pi}\be(s_{a}) \\ & &\quad\sum_{\de,\ep}
\int \frac{d\Omega_{da}}{4\pi}N^{*}_{\de\ep;\be\ga}(s_{a}
,\cos\th_{bd})\cdot M_{\al\de\ep;\rho}(s_{a},\cos\th_{da}) \nonumber \eeqq
where we have introduced the shorthand notation
\beq \be(s)\equiv\sqrt{\frac{s-4}{s}}. \eeq

We note that some of the kinematical variables are not
well behaved in the decay region; we shall
supplement a correct analytic interpretation for them below.

Adopting the normalizations from Ref. \cite{GLPP}, we continue
with the isospin decomposition of the invariant
pion-pion scattering matrix element
\beqq \label{3.7}
N_{\al\be;\ga\de}(s_{a},s_{b},s_{c})
&=&\frac{1}{3}(T^{0}(s_{a},s_{b},s_{c})-T^{2}
(s_{a},s_{b},s_{c}))\de_{\al\be}\de_{\ga\de} \nonumber \\
&+&\frac{1}{2}(T^{2}(s_{a},s_{b},s_{c})+T^{1}
(s_{a},s_{b},s_{c}))\de_{\al\ga}\de_{\be\de} \\
&+&\frac{1}{2}(T^{2}(s_{a},s_{b},s_{c})-T^{1}
(s_{a},s_{b},s_{c}))\de_{\al\de}\de_{\be\ga} \nonumber \eeqq
where the $T^{I}$ are the isospin amplitudes
to isospin $I$. In our frame of reference
they depend on two variables only, the
total invariant mass of the pions and the
intermediate scattering angle $\th_{bd}$. The corresponding isospin
decomposition of the decay amplitude has already
been given in eq. (\ref{2.2}).

We now insert  eq. (\ref{3.7}) into eq. (\ref{3.5}) and use
eq. (\ref{2.2}) This yields for the
discontinuity of $M$ in $s_{a}$
\beqq & &\disc_{|_{\mbox{a}}} M^{1}_{\al\be\ga;\rho}=\frac{1}{32\pi}
\be(s_{a}) \int\frac{d\Omega_{da}}{4\pi} \nonumber \\
& &\quad\quad[
\de_{\be\ga}\de_{\al\rho}\{T^{0*}(s_{a},s'_{d},s'_{e})F(s_{a},s_{d},s_{e})
\nonumber \\ & &\quad\quad\quad
+\frac{1}{3}(T^{0*}(s_{a},s'_{d},s'_{e})-T^{2*}(s_{a},s'_{d},s'_{e}))
F(s_{e},s_{a},s_{d}) \nonumber \\
& &\quad\quad\quad
+\frac{1}{3}(T^{0*}(s_{a},s'_{d},s'_{e})-T^{2*}(s_{a},s'_{d},s'_{e}))
F(s_{d},s_{e},s_{a})\} \\ & &\quad\quad
+\de_{\ga\al}\de_{\be\rho}\{\frac{1}{2}(T^{2*}(s_{a},s'_{d},s'_{e})+T^{1*}
(s_{a},s'_{d},s'_{e}))F(s_{d},s_{e},s_{a}) \nonumber \\ & &\quad\quad\quad
+\frac{1}{2}(T^{2*}(s_{a},s'_{d},s'_{e})-T^{1*}(s_{a},s'_{d},s'_{e}))
F(s_{e},s_{a},s_{d})\} \nonumber \\ & &\quad\quad
+\de_{\al\be}\de_{\ga\rho}\{\frac{1}{2}(T^{2*}(s_{a},s'_{d},s'_{e})+T^{1*}
(s_{a},s'_{d},s'_{e}))F(s_{e},s_{a},s_{d}) \nonumber \\ & &\quad\quad\quad
+\frac{1}{2}(T^{2*}(s_{a},s'_{d},s'_{e})-T^{1*}(s_{a},s'_{d},s'_{e}))
F(s_{d},s_{e},s_{a})\}]. \nonumber \eeqq

Next, we use this result and the two analogous ones for the
discontinuities in $s_{b}$ and $s_{c}$ in the dispersion
representation eq. (\ref{3.1}) for the full amplitude.
Comparing the coefficients in the different isospin
channels we obtain three dispersion relations for the
scalar functions $F$ \cite{KT60}:
\beqq \label{3.9}  F(s_{a},s_{b},s_{c})&=&\frac{1}{\pi}
\intgr\frac{ds'_{a}}{s'_{a}-s_{a}-i\ep}
\,\frac{1}{32\pi}\be(s'_{a})\int\frac{d\Omega_{da}}{4\pi} \nonumber \\
& &\quad\quad\{T^{0*}(s'_{a},s'_{d},s'_{e})F(s'_{a},s_{d},s_{e})
\nonumber \\
& &\quad\quad+\frac{1}{3}(T^{0*}(s'_{a},s'_{d},s'_{e})
-T^{2*}(s'_{a},s'_{d},s'_{e}))F(s_{d},s_{e},s'_{a}) \nonumber \\
& &\quad\quad+\frac{1}{3}(T^{0*}(s'_{a},s'_{d},s'_{e})
-T^{2*}(s'_{a},s'_{d},s'_{e}))F(s_{e},s'_{a},s_{d})\} \nonumber \\
&+&\frac{1}{\pi}\intgr\frac{ds'_{b}}{s'_{b}-s_{b}-i\ep}
\,\frac{1}{32\pi}\be(s'_{b})\int\frac{d\Omega_{db}}{4\pi} \\
& &\quad\quad\{\frac{1}{2}(T^{2*}(s'_{b},s'_{d},s'_{e})+T^{1*}
(s'_{b},s'_{d},s'_{e}))F(s_{e},s'_{b},s_{d}) \nonumber \\
& &\quad\quad+\frac{1}{2}(T^{2*}(s'_{b},s'_{d},s'_{e})
-T^{1*}(s'_{b},s'_{d},s'_{e}))
F(s_{d},s_{e},s'_{b})\} \nonumber \\
&+&\frac{1}{\pi}\intgr\frac{ds'_{c}}{s'_{c}-s_{c}-i\ep}
\,\frac{1}{32\pi}\be(s'_{c})\int\frac{d\Omega_{dc}}{4\pi} \nonumber \\
& &\quad\quad\{\frac{1}{2}(T^{2*}(s'_{d},s'_{e},s'_{c})+T^{1*}
(s'_{c},s'_{d},s'_{e}))F(s_{d},s_{e},s'_{c}) \nonumber \\
& &\quad\quad+\frac{1}{2}(T^{2*}(s'_{c},s'_{d},s'_{e})
-T^{1*}(s'_{c},s'_{d},s'_{e}))
F(s_{e},s'_{c},s_{d})\}. \nonumber \eeqq
The invariant variables $s_{d},s_{e},s'_{d},s'_{e}$
must be expressed in terms of $s_{a},s_{b},s_{c}$.
The other two scalar dispersion representations are obtained from
eq. (\ref{3.9}) by
permutation of $a,b,c$.

To evaluate further this dispersion relation we retain
only the $s$- and $p$-wave contributions to the rescattering.
 The relevant partial wave expansions of the pion
scattering amplitudes $T^{I}$ for fixed isospin $I$
are given as usual by \cite{GLPP}
\beq \label{3.12} T^{I}(s,\cos\th)
=\frac{32\pi}{\be(s)}\sum_{0}^{\infty}(2l+1)P_{l}
(\cos\th)f_{l}^{I}(s) \eeq
where $s$ is the total invariant mass and $\th$
the scattering angle in the c.m. system of the
colliding pions. In the elastic region $4<s<16$
the partial wave amplitudes are parametrized by
real phase shifts as
\beq f_{l}^{I}(s)=e^{i\de_{l}^{I}(s)}\sin\de_{l}^{I}(s). \eeq
The lowest contributions for $I=0,2$ are $s$-waves
\beqq \label{3.13} T^{0}(s)&=&\frac{32\pi}{\be(s)}f_{0}^{0}(s) \\
T^{2}(s)&=&\frac{32\pi}{\be(s)}f_{0}^{2}(s) \nonumber \eeqq
while for $I=1$ it is a $p$-wave
\beq T^{1}(s,\cos\th)
=\frac{32\pi}{\be(s)}3\cos\th\cdot f_{1}^{1}(s). \eeq
The factors $\frac{1}{32\pi}\be$ in (\ref{3.9}) just cancel
now against the corresponding factors in the partial
wave expansion of $T^{I}$ and we are left with two types
of angle integrals.

In the case of $s$-wave contributions the pion-pion rescattering
is isotropic. For $I=0,2$ we have thus angle integrations of the type
\cite{KT60}
\beqq {\overline F}(s_{a})&\equiv&
\int\frac{d\Omega_{da}}{4\pi}F(s_{a},s_{d},s_{e}) \\
{\tilde F}(s_{a})&\equiv&\int
\frac{d\Omega_{da}}{4\pi}F(s_{d},s_{e},s_{a}) \nonumber \eeqq
where $s_{d},s_{e}$ are expressed through $s_{a}$
and $\cos\th_{da}$ as
\beqq s_{e}&=&3s_{0}-s_{a}-s_{d} \nonumber \\
s_{d}&=&\frac{3s_{0}-s_{a}}{2}+K(s_{a})\cdot\cos\th_{da} \\
d\Omega_{da}&=&d\va\cdot d\cos\th_{da}. \nonumber \eeqq
Here we introduce the Kacser function \cite{KAK}
\beq \label{3.16} K(s)\equiv\sqrt{\frac{(s-4)(s-(m-1)^{2})
(s-(m+1)^{2})}{4s}}. \eeq
Its cuts and limits in the complex $s$-plane will
be discussed below.

We may now recast the integrations above as
integrations over $s_{d}$ and obtain
\beq \label{3.17}
{\overline F}(s_{a})=\frac{1}{2K(s_{a})}
\int_{s_{d-}(s_{a})}^{s_{d+}(s_{a})}
ds_{d} \,F(s_{a},s_{d},3s_{0}-s_{a}-s_{d}) \eeq
and in an analogous way
\beq \label{3.18} {\tilde F}(s_{a})=\frac{1}{2K(s_{a})}
\int_{s_{d-}(s_{a})}^{s_{d+}(s_{a})} ds_{d}
\,F(s_{d},3s_{0}-s_{a}-s_{d},s_{a}) \eeq
where
\beq s_{d+}(s)\equiv\frac{3s_{0}-s}{2}+K(s),\quad s_{d-}(s)
\equiv\frac{3s_{0}-s}{2}-K(s). \eeq
For later use we note a symmetry property of the integral
${\tilde F}(s_{a})$
\beq {\tilde F}(s_{a})=\frac{1}{2K(s_{a})}
\int_{s_{d-}(s_{a})}^{s_{d+}(s_{a})} ds_{d}
F(3s_{0}-s_{a}-s_{d},s_{a},s_{d}) \eeq
which follows immediately from the Bose
symmetry properties of $F$ itself and a
change of variable $s_{d}\ar 3s_{0}-s_{a}-s_{d}$.

The $p$-wave contribution for $I=1$ leads to the more complicated integral
\beq I(s_{a},s_{b},s_{c})\equiv\int\frac{d\Omega_{da}}{4\pi}\cos\th_{bd}
F(s_{e},s_{a},s_{d}). \eeq
In order to perform the angle integration
explicitly and to get rid of the two angles
we invoke a partial wave expansion for $F$
\beq F(s_{e},s_{a},s_{d})
=\sum_{0}^{\infty}(2l+1)P_{l}(\cos\th_{da})F_{l}^{s_{d}}(s_{a}).
\nonumber  \eeq
Making use of well-known properties of the
Legendre polynomials we obtain
\beq I(s_{a},s_{b},s_{c})=\cos\th_{ba}
\cdot F_{1}^{s_{d}}(s_{a}). \eeq
The $l=1$ projection of $F$ is next expressed as
an integral over $s_{d}$
\beqq F_{1}^{s_{d}}(s_{a})&=&\frac{1}{4K^{2}(s_{a})}
\int_{s_{d-}(s_{a})}^{s_{d+}(s_{a})} ds_{d} \,(2s_{d}+s_{a}-3s_{0}) \\
& &\quad\quad \cdot F(3s_{0}-s_{a}-s_{d},s_{a},s_{d})
\nonumber \eeqq
and $\cos\th_{ba}$ is recast as
\beq \cos\th_{ba}=\frac{s_{b}-s_{c}}{2K(s_{a})}. \eeq
Plugging all together we obtain the desired expression
\beqq I(s_{a},s_{b},s_{c})&=&
\frac{s_{b}-s_{c}}{8K^{3}(s_{a})}
\int_{s_{d-}(s_{a})}^{s_{d+}(s_{a})} ds_{d} \,(2s_{d}+s_{a}-3s_{0}) \\
& &\quad\quad \cdot F(3s_{0}-s_{a}-s_{d},s_{a},s_{d}). \nonumber \eeqq
Let us note for later use a symmetry property of $I(s_{a},s_{b},s_{c})$
\beqq I(s_{a},s_{b},s_{c})&=&
-\frac{s_{b}-s_{c}}{8K^{3}(s_{a})}
\int_{s_{d-}(s_{a})}^{s_{d+}(s_{a})} ds_{d} \,(2s_{d}+s_{a}-3s_{0}) \\
& &\quad\quad \cdot F(s_{d},3s_{0}-s_{a}-s_{d},s_{a}) \nonumber \eeqq
following from the Bose symmetry properties of $F$ and a
change of variable $s_{d}\ar 3s_{0}-s_{a}-s_{d}$. Finally we
decompose $I(s_{a},s_{b},s_{c})$ into
\beq I(s_{a},s_{b},s_{c})=\frac{s_{b}-s_{c}}{3}{\hat F}(s_{a}), \eeq
defining the function of one variable
\beqq \label{3.29}{\hat F}(s_{a})&\equiv&
\frac{3}{8K^{3}(s_{a})}\int_{s_{d-}(s_{a})}^{s_{d+}(s_{a})} ds_{d}
\,(2s_{d}+s_{a}-3s_{0}) \\ & &\quad\quad
\cdot F(3s_{0}-s_{a}-s_{d},s_{a},s_{d}).
\nonumber \eeqq
This will allow us to rewrite the integral equation
for $F(s_{a},s_{b},s_{c})$ as an equivalent set of
coupled integral equations for functions of one variable only.

Note that the projections as given in eqs. (\ref{3.17}),
(\ref{3.18}) and (\ref{3.29}) are defined only for values of
$s_{a}$ in the physical decay region $4<s_{a}<(m-1)^{2}$.
Outside this range the argument of the square root in the
Kacser function eq. (\ref{3.16}) becomes negative and we
have to give this funcion a meaning by appropriate analytic
continuation. This was studied by Bronzan and Kacser \cite{BK,KAK}
whom we follow closely.
Based on comparison with explicit expressions for the
projection operation in perturbation theory and demanding
that the general definition of the projection should reduce
to these expressions, they find that the naive integrals
have to be replaced by contour integrals in the complex
$s_{d}$-plane. The respective paths joining $s_{d-}(s_{a})$
with $s_{d+}(s_{a})$ must avoid the real axis
for $4<s_{d}<\infty$ as the amplitude $F$ has a cut there.
Where necessary, $s_{d-}(s_{a})$ and $s_{d+}(s_{a})$ are
taken infinitesimally above or below the real axis according
to the prescription obtained by replacing
$m^{2}\ar m^{2}+i\de$, $\de\ar 0+$ for real $s_{a}$.
Hence, we redefine the projections as the contour integrals
\beqq {\overline F}(s_{a})&=&\frac{1}{2K(s_{a})}
\oint_{s_{d-}(s_{a})}^{s_{d+}(s_{a})}
ds_{d} \,F(s_{a},s_{d},3s_{0}-s_{a}-s_{d}), \nonumber \\
{\tilde F}(s_{a})&=&\frac{1}{2K(s_{a})}
\oint_{s_{d-}(s_{a})}^{s_{d+}(s_{a})} ds_{d}
\,F(s_{d},3s_{0}-s_{a}-s_{d},s_{a}), \\
{\hat F}(s_{a})&=&\frac{3}{8K^{3}(s_{a})}
\oint_{s_{d-}(s_{a})}^{s_{d+}(s_{a})}
s_{d} \,(2s_{d}+s_{a}-3s_{0}) \nonumber \\ & &\quad\quad
\cdot F(3s_{0}-s_{a}-s_{d},s_{a},s_{d}) \nonumber \eeqq
and introduce the function $H$
\beq H(s)\equiv\sqrt{\frac{|(s-4)(s-(m-1)^{2})(s-(m+1)^{2})|}{|4s|}} \eeq
which is well defined for all $s$ different from zero and allows us to express
the necessary
analytic continuation of the Kacser function $K$ in a simple way.

Next we turn to the description of the different paths
in the $s_{d}$-plane belonging to the four cases we have to discuss
if evaluating the aforementioned definition of the projection operation.
We must distinguish four cases.

\noindent i) $4<s_{a}<\frac{m^{2}-1}{2}$: as $s_{a}$ is
in the physical decay region $K(s_{a})$ coincides with $H(s_{a})$.
(see Fig. 1a). The end points of the contour are lying
infinitesimally above the real axis and are found to be
\beqq s_{d+}(s_{a})&=&\frac{3s_{0}-s_{a}}{2}+H(s_{a})+i\de, \\
s_{d-}(s_{a})&=&\frac{3s_{0}-s_{a}}{2}-H(s_{a})+i\de. \nonumber \eeqq

\noindent ii) $\frac{m^{2}-1}{2}<s_{a}<(m-1)^{2}$: $s_{a}$ is
still in the physical decay region so that $K(s_{a})$ coincides with
$H(s_{a})$ again.
The point $s_{a}=\frac{m^{2}-1}{2}$ is important
because when $s_{a}$ goes through it, the endpoint
$s_{d-}(s_{a})$ continuously turns around the
beginning of the cut at $4$ and remains then
infinitesimally below the real axis (see Fig. 1b)
such that only the other endpoint $s_{d+}(s_{a})$ of the contour
remains infinitesimally above the real axis. The two end points are
now found to be
\beqq s_{d+}(s_{a})&=&\frac{3s_{0}-s_{a}}{2}+H(s_{a})+i\de, \\
s_{d-}(s_{a})&=&\frac{3s_{0}-s_{a}}{2}-H(s_{a})-i\de. \eeqq

\noindent iii) $(m-1)^{2}<s_{a}<(m+1)^{2}$:
as $s_{a}$ has left the physical decay region
$K(s_{a})$ must be analytically continued and
is defined to be $K(s_{a})\equiv iH(s_{a})$.
The endpoint $s_{d-}(s_{a})$ turns now smoothly into the
 half plane below the real axis whereas the other
endpoint $s_{d+}(s_{a})$ turns smoothly into the
half plane above the real axis in a symmetric fashion such that
\beq \label{3.35}s_{d+}(s_{a})=\frac{3s_{0}-s_{a}}{2}+iH(s_{a}),
\quad s_{d-}(s_{a})=\frac{3s_{0}-s_{a}}{2}-iH(s_{a}). \eeq
The most adequate contour for numerical purposes joining
those two points follows just the path the end points
move along if $s_{a}$ goes from $(m-1)^{2}$ to
$(m+1)^{2}$ (see Fig. 1c) and is thus parametrized
by $s$ itself as shown in eqns.(\ref{3.35}).

\noindent iv) $(m+1)^{2}<s_{a}$: as $s_{a}$ is
out of the physical decay region $K(s_{a})$ must
again be analytically continued and is defined
to be $K(s_{a})\equiv -H(s_{a})$ (see Fig. 1d).
The end points of the contour are lying in this case on the real axis
\beq s_{d+}(s_{a})=\frac{3s_{0}-s_{a}}{2}-H(s_{a}),\quad s_{d-}(s_{a})
=\frac{3s_{0}-s_{a}}{2}+H(s_{a}). \eeq

Note that at $s_{a}=(m-1)^{2}$ the projections of a
function with a cut have a singularity because
$K$ vanishes at this point whereas the
integration contour still has finite length and
the integral over the imaginary part of the function
along the contour does not vanish. The actual form of this
singularity for ${\overline F}(s_{a})$ around
$s_{a}=(m-1)^{2}$ is then given by
\beq \label{3.37} {\overline
F}(s_{a})\sim\frac{1}{((m-1)^{2}-s_{a})^{\frac{1}{2}}}
\cdot\mbox{const.}
\eeq
and for ${\hat F}(s_{a})$ by
\beq \label{3.38} {\hat F}(s_{a})\sim\frac{1}{((m-1)^{2}-s_{a})^{\frac{3}{2}}}
\cdot\mbox{const.}
\eeq
where the constant is in fact the value of the contour
integral for $s_{a}=(m-1)^{2}$. We will use this form
later to analyse the consequences of this pole in the
projection prescription.

The projection operations occuring in the eqs. (\ref{3.17}),
(\ref{3.18}) and (\ref{3.29}) are now given by well defined
contour integrals and we use them in the further
evaluation of the integral equation (\ref{3.9}) for $F$.
We thus insert the partial wave expansions of $T^{I}$ in
the expression
 (\ref{3.9}) for the scalar function $F$. Using the
definitions (\ref{3.17}), (\ref{3.18}) and (\ref{3.29}) for
the different projections and including the subtraction polynomial as
required by the discussion in section 4, we may recast $F$
in the following way
\beqq
\label{3.39}F(s_{a},s_{b},s_{c})
&=&P(s_{a},s_{b},s_{c})+U(s_{a})+V(s_{b})+V(s_{c}) \\
&+&(s_{c}-s_{a})\cdot W(s_{b})+(s_{b}-s_{a})\cdot W(s_{c}) \nonumber \eeqq
where we defined the functions of one variable $U,V,W$
\beqq \label{3.40} U(s_{a})&=&\frac{\prod_{i=1}^{3}
(s_{a}-s_{i})}
{\pi}\intgr\frac{ds'_{a}}{\prod_{j=1}^{3}
(s'_{a}-s_{j})(s'_{a}-s_{a}-i\ep)} \\
& & \quad\cdot\{f^{0*}_{0}(s'_{a}){\overline F}(s'_{a})
+\frac{2}{3}(f^{0*}_{0}(s'_{a})-f^{2*}_{0}(s'_{a})){\tilde F}(s'_{a})\},
\nonumber \eeqq
\beqq \label{3.41}  V(s_{a})&=&\frac{\prod_{i=1}^{3}
(s_{a}-s_{i})}{\pi}
\intgr\frac{ds'_{a}}{\prod_{j=1}^{3}(s'_{a}-s_{j})
(s'_{a}-s_{a}-i\ep)} \\ & &
\quad\quad\quad\quad\quad\quad\quad
\cdot f^{2*}_{0}(s'_{a}){\tilde F}(s'_{a})  \nonumber \eeqq
and
\beqq \label{3.42} W(s_{a})&=&\frac{(s_{a}-s_{1})(s_{a}-s_{2})}{3\pi}
\intgr\frac{ds'_{a}}{(s'_{a}-s_{1})(s'_{a}-s_{2})(s'_{a}-s_{a}-i\ep)}
\nonumber \\ & &
\quad\quad\quad\quad\quad\quad\quad
\cdot f^{1*}_{1}(s'_{a}){\hat F}(s'_{a})  \\
&+&\mbox{perm.},\nonumber \eeqq
where perm. denotes the symmetrization of $W$ in
$s_1,s_2,s_3$. Note that, unlike $U$ and $V$, the
function $W$ is subtracted only twice. As ChPT yields
the amplitude $F$ only to the accuracy $p^{4}$ and as
$W$ is multiplied by factors of $O(p^{2})$ we can not
make three subtractions here. On the other hand,
convergence of the integral is still
ensured as the $p$-wave projection
brings an extra factor of $\frac{1}{s'_{a}}$ coming from
the Kacser function $K$ in the integrand.

Numerically it is easier to deal with three integral
equations for three functions of one variable than with
one such equation for a function of three variables. We
thus have to transform the content of the subtracted
dispersion representation eq. (\ref{3.9}) for $F$ by
taking its $s$- and $p$-wave projections. Note that all
projections of functions of one variable are either
trivial or may be expressed as a "bar-projection"
(see eq. (\ref{3.17}), where
each argument $s_{c}$ is understood to be expressed
by the integration variable $s_{b}$ and constant terms by
\beq s_{c}=3s_{0}-s_{a}-s_{b}. \eeq
The projections become now
\beqq \label{3.43} {\overline F}(s_{a})
&=&{\overline P}(s_{a})+U(s_{a})+{\overline V_{b}}(s_{a})+
{\overline V_{c}}(s_{a}) \nonumber \\
&+&{\overline{(s_{c}-s_{a})W_{b}}}\,(s_{a})+
{\overline{(s_{b}-s_{a})W_{c}}}\,(s_{a}), \nonumber \\
{\tilde F}(s_{a})&=&{\tilde P}(s_{a})+
{\overline U_{b}}(s_{a})+{\overline V_{c}}(s_{a}) +V(s_{a})
\nonumber \\
&+&{\overline{(s_{a}-s_{b})W_{c}}}\,(s_{a}) \nonumber \\
{\hat F}(s_{a})&=&{\hat P}(s_{a})
+\frac{3}{4K^{2}(s_{a})}
\{{\overline{(s_{b}-s_{c})U_{c}}}\,(s_{a})
+{\overline{(s_{b}-s_{c})V_{b}}}\,(s_{a}) \nonumber \\
&+&{\overline{(s_{b}-s_{c})^{2}}}\cdot W(s_{a})+
{\overline{(s_{b}-s_{c})(s_{a}-s_{c})W_{b}}}\,(s_{a})\}
\nonumber \eeqq
where the subscripts $b,c$ denote the argument of
the respective function under the projection integral.
 Note that certain terms vanish because
${\overline{s_{b}-s_{c}}}=0$.
We next change
arguments of the functions above to $s_{b}$ and omit in
the following the subscripts making use of the relations
\beqq {\overline A_{c}}(s_{a})&=&{\overline A_{b}}(s_{a}) \nonumber \\
{\overline{s_{b}\,A_{c}}}(s_{a})&=&{\overline{s_{c}\,A_{b}}}(s_{a}) \\
{\overline{s_{c}\,A_{c}}}(s_{a})&=&{\overline{s_{b}\,A_{b}}}(s_{a})
\nonumber \eeqq
where $A$ denotes any of the functions $U,V,W$.
Eqns.(\ref{3.43}) are thus rewritten as
\beqq \label{3.45}
{\overline F}(s_{a})&=&{\overline P}(s_{a})+U(s_{a})+
2{\overline V}(s_{a}) \nonumber \\
&+&2{\overline{(s_{c}-s_{a})W}}\,(s_{a}), \nonumber \\
{\tilde F}(s_{a})&=&{\tilde P}(s_{a})+
{\overline U}(s_{a})+V(s_{a})+{\overline V}(s_{a}) \nonumber \\
&+&{\overline{(s_{a}-s_{c})W}}\,(s_{a}), \nonumber \\
{\hat F}(s_{a})&=&{\hat P}(s_{a})
+\frac{3}{4K^{2}(s_{a})}\{{\overline{(s_{c}-s_{b})U}}\,(s_{a})
+{\overline{(s_{b}-s_{c})V}}\,(s_{a}) \nonumber \\
&+&{\overline{(s_{b}-s_{c})^{2}}}\cdot W(s_{a})+
{\overline{(s_{b}-s_{c})
(s_{a}-s_{c})W}}\,(s_{a})\}. \nonumber \eeqq

Finally, we insert eqs. (\ref{3.45}) for
the projections expressed in terms of $U,V,W$ into
the defining eq. (\ref{3.40}) for $U$, eq. (\ref{3.41} for
$V$ and (\ref{3.42}) for $W$. This yields the coupled system
\beqq \label{3.47}
U(s_{a})&=&\frac{\prod_{i=1}^{3}(s_{a}-s_{i})}{\pi}
\intgr\frac{ds'_{a}}{\prod_{j=1}^{3}
(s'_{a}-s_{j})(s'_{a}-s_{a}-i\ep)} \nonumber \\
& & \quad\cdot\{f^{0*}_{0}(s'_{a})\,[{\overline P}
(s'_{a})+\frac{2}{3}{\tilde P}(s'_{a})+U(s'_{a})+
\frac{2}{3}{\overline U}(s'_{a}) \nonumber \\
& & \quad\quad+\frac{2}{3}V(s'_{a})+\frac{8}{3}{\overline V}(s'_{a})
+\frac{4}{3}{\overline{(s_{c}-s_{a})W}}\,(s'_{a})] \\
& & \quad-f^{2*}_{0}(s'_{a})\,[\frac{2}{3}
{\tilde P}(s'_{a})+\frac{2}{3}
{\overline U}(s'_{a})+\frac{2}{3}V(s'_{a}) \nonumber \\
& & \quad\quad+\frac{2}{3}{\overline V}(s'_{a}) +
\frac{2}{3}{\overline{(s_{a}-s_{c})W}}\,(s'_{a})]\}, \nonumber \eeqq
\beqq \label{3.48}
V(s_{a})&=&\frac{\prod_{i=1}^{3}(s_{a}-s_{i})}
{\pi}\intgr\frac{ds'_{a}}
{\prod_{j=1}^{3}(s'_{a}-s_{j})(s'_{a}-s_{a}-i\ep)} \nonumber \\
& & \quad\cdot f^{2*}_{0}(s'_{a})\,[{\tilde P}
(s'_{a})+{\overline U}(s'_{a})+V(s'_{a}) \\
& & \quad\quad+{\overline V}(s'_{a})+{\overline{(s_{a}-s_{c})W}}\,(s'_{a})]
\nonumber \eeqq
and
\beqq \label{3.49} & &
W(s_{a})=\frac{(s_{a}-s_{1})(s_{a}-s_{2})}{3\pi}
\intgr\frac{ds'_{a}}{(s'_{a}-s_{1})(s'_{a}-s_{2})
(s'_{a}-s_{a}-i\ep)} \nonumber \\ & &
\quad\quad\cdot f^{1*}_{1}(s'_{a})\,[{\hat P}(s'_{a})
+\frac{3}{4K^{2}(s'_{a})}\{{\overline{(s_{c}-s_{b})U}}\,(s'_{a})
+{\overline{(s_{b}-s_{c})V}}\,(s'_{a}) \\
& & \quad\quad +{\overline{(s_{b}-s_{c})^{2}}}
\cdot W(s'_{a})+{\overline{(s_{b}-s_{c})
(s_{a}-s_{c})W}}\,(s'_{a})\}] \nonumber \\
&+&\mbox{perm.} \nonumber \eeqq
These are our final expressions which generalize the
Khuri-Treiman equations.
As we mentioned above there are in fact singularities
in the projected functions occuring under the
integrals. The question of whether this leads to
problems in the reconstruction of the amplitude
by the dispersion integrals will be analysed in section 5.

\Section{Subtractions}
The generalized Khuri-Treiman equations in
(\ref{3.39}-\ref{3.42}) have been formulated
with three subtractions, because we will use them in this
form. However, as usual in dispersion relation techniques, the
number of subtractions is to a large  extent arbitrary. The minimal
number of subtractions is determined by the high energy behaviour
of the amplitude. If this is known, one is of course still free to take
more subtractions than necessary for convergence.
With fewer subtractions, the dispersion integral is more weakly convergent
and depends more strongly on
the poorly known high energy behaviour of the scattering phase.
Conversely, if many subtractions
are used  the integral is better behaved;
however then we need a larger number of subtraction
constants which in general are less well known.
Thus, one must keep a delicate balance in order to optimally
use the available data and their uncertainties.

The asymptotic behaviour of the $\eta\rightarrow 3\pi$ amplitude may
be indicated by Regge phenomenology. Since there is no pomeron exchange
in $\eta\pi\rightarrow\pi\pi$, the leading Regge trajectory
 is the one associated with the $\rho$;
we therefore expect an asymptotic behaviour $A \sim s_a^{1/2}$ ($s_b=0$)
for the $I=1$ component of the amplitude in the t-channel.
The $I=0,2$ components are not affected
by $\rho$-exchange and are expected to approach a constant
asymptotically. In order to ensure convergence, it is thus sufficient
to subtract the dispersion relation once.

However, our calculational scheme of dealing with $\pi\pi$ final state
interactions
relies on elastic unitarity. In the $I=0$ s-wave channel the phase shift
exhibits large inelasticities above 1 GeV. Therefore it is
desirable that
this region is unimportant in the evaluation of the
dispersion integrals.
It turns out that this requirement implies at least two subtractions.
We will discuss the influence of the phase
shifts above 1 GeV on our results in sect. 6.

Besides the final state interactions we are attempting to control, there
are also mass corrections to the $\eta\rightarrow 3\pi$ amplitude. These
are also known at the one-loop level where the total
amplitude is \cite{GL85c}
\begin{eqnarray}
A^{\rm 1-loop}(s_a,s_b,s_c) 	=
-
{Q^{-2}\frac
{m_K^2}{m_\pi^2}(m_K^2-m_\pi^2)
\over 3 \sqrt{3} F_\pi^2} \times \nonumber \\
\left\{ T(s_a)+
{\cal U}(s_a,s_b,s_c) 	+{\cal V}(s_a)+{\cal W}(s_a,s_b,s_c) 	\right\},
\label{A1-loop}
\end{eqnarray}
where according to the discussion in section 1, the normalization
of the amplitude is given by the first numerator rather than by
\begin{equation}
m_1^2=m_{K^0}^2-m_{K^+}^2-m_{\pi^0}^2+m_{\pi^+}^2
\end{equation}
which would follow if Dashen's theorem is used to
determine the electromagnetic mass difference of the kaons.
The leading order expression $T(s_a)$ is given in eq. (\ref{amp})
and the other term will be discussed below.
 ${\cal U}(s_a,s_b,s_c) $ denotes the unitarity
corrections and ${\cal V}(s_a)$, ${\cal W}(s_a,s_b,s_c) $ are polynomials
in $s_a,s_b,s_c$  resulting from
tadpole and tree graphs at order $p^4$ in the low energy expansion.
${\cal U}(s_a,s_b,s_c) $
contains the two-pion $I=0,1$ and $2$ final state interactions. These
rescattering
graphs are primarily responsible for a large enhancement of
the amplitude at
next-to-leading order. However, as discussed in great detail in Ref.
\cite{GL85c}, only the sum of all terms in eq. (\ref{A1-loop}) is a meaningful
quantity: the magnitude of the final state interactions of the pions depends
on the manner in which the unitarity corrections are split off from the rest.

In this work we use the ChPT one-loop amplitude
to obtain an improved
estimate for the subtraction polynomial entering the generalized
Khuri-Treiman equations.
There are three important points here:
i) As mentioned above the one-loop amplitude
contains
already some final state interactions. In order
to avoid double counting, these final state interactions have to be
incorporated in a well defined manner. ii) The choice of the subtraction
points: In principle, a dispersion relation can be subtracted at
any point. However, since the amplitude is approximated by a
polynomial near these points, they should lie as far as possible
from kinematical singularities. Furthermore, we expect that the low energy
expansion is more reliable, if the invariant mass of the interacting pion pair
is as small as possible at the subtraction points.
However, as soon as $I=2$ (and $I=1$) final state interactions are
included, the
problem depends on two variables $s_a$ and $s_b$, i.e. the pions interact
also in the
$s_b$- and $s_c$ -channels. This is reflected by the fact that there
are three dispersion integrals. They should all be subtracted
at small momenta, but there is no point in the Mandelstam plane
where all three variables
$s_a$, $s_b$ and $s_c$ are small.
iii) The subtraction constants as determined from the existing one-
loop calculation will have uncertainties due to yet
unknown higher order effects. These uncertainties can be estimated by
comparing the one-loop amplitude with the leading order term. A precise
description of our estimate will be given in section 6.

The solution to the first two problems is straightforward and motivated by the
following observation. In chiral perturbation theory,
the imaginary part of the $\eta\rightarrow 3 \pi$ one-loop amplitude
is generated by the graphs where $\pi\pi$, $\pi\eta$, $K \bar K$ or
$\eta \eta$ rescatter.
They
are easily obtained from two-body unitarity provided the lowest order vertices
are inserted in the unitarity relation. Given the absorptive parts, the real
part at order $p^4$ can be reconstructed from a dispersion representation
up to a second order polynomial in $s_a, s_b, s_c$. Due to the work of Gasser
and
Leutwyler \cite{GL85c}, this polynomial is known in terms of $m_\pi$, $m_K$,
$m_\eta$,
$F_\pi$, $F_K$ and the low-energy constant $L_3$.

The Khuri-Treiman equations are of course more general than this
dispersion representation of the $\eta\rightarrow 3 \pi$ amplitude to order
$p^4$. However, they can be matched onto the ChPT one-loop amplitude in the
following sense. Consider an iterative solution of eqs.
(\ref{3.39} - \ref{3.42}).
In particular, use
the current algebra result, $F(s_a,s_b,s_c)^{(0)}=T(s_a)$, as a first
approximation to be inserted on
the right hand
side of eq. (\ref{3.40})
and the partial waves of the $\pi\pi$ scattering amplitude as
given by their low energy expansion, i.e.
\begin{equation}
f_l^I(s)=e^{i\delta_l^I(s)} \sin \delta_l^I(s)
\approx \delta_l^{I,{\rm ChPT}}(s)
\end{equation}
where $\delta_l^{I, {\rm ChPT}}(s)$ are the expressions for
the phaseshifts at
leading order ChPT given in Ref. \cite{GL85c}. By construction,
the first step
of this iteration yields exactly the imaginary parts of the ChPT one-loop
amplitude which are due to $\pi\pi$ intermediate states
\footnote{The absorptive parts due to $\pi\eta$, $K \bar K$ or $\eta \eta$
intermediate
states were not included in the calculation of the imaginary part.
Because of
their high thresholds, we expect a small effect and will neglect
them.}.
Moreover, the dispersive part of the Khuri-Treiman equations contains also
{\it all} $\pi\pi$-threshold effects present in the ChPT one-loop
amplitude.
Therefore, after the first iteration step as described above, we have
reproduced all singularities close to the physical region.
We note already at this point that three subtractions will be necessary
in order
to make this procedure well defined.

In the low energy
region, the remainder of the ChPT one-loop amplitude can be
expanded to a very
good accuracy into a polynomial of second order in $s_a,s_b,s_c$.
This last step then
determines the subtraction polynomial of the generalized Khuri-Treiman
equations.

Before turning to the details of our subtraction prescription,
we give a more
complete discussion of the structure of the one-loop
amplitude in Chiral Perturbation Theory.

\subsection{Structure of the ChPT one-loop amplitude}

We work with the reduced amplitude $\bar A$ defined by
\begin{equation}
A^{\rm 1-loop}(s_a,s_b,s_c) 	=-
{Q^{-2}\frac
{m_K^2}{m_\pi^2}(m_K^2-m_\pi^2)
 \over 3 \sqrt{3} F_\pi^2} \bar A(s_a,s_b,s_c)
\end{equation}
where $A^{\rm 1-loop}(s_a,s_b,s_c)$ is given in eq. (\ref{A1-loop}). $T$,
$\cal V$ and $\cal W$
are polynomials to second order in $s_a,s_b,s_c$. The aim
is to rewrite $A^{\rm 1-loop}(s_a,s_b,s_c) 	$ as a suitable
dispersion integral. The singularities of the amplitude
$\bar A$ are contained in the unitarity corrections
${\cal U}(s_a,s_b,s_c) 	$. We may split them
up further by writing
\begin{equation}
{\cal U}(s_a,s_b,s_c) 	={\cal U}^{\rm disp}_{\pi\pi}(s_a,s_b,s_c) +{\cal U}
^{\rm rem}(s_a,s_b,s_c) ,
\end{equation}
with
\begin{eqnarray}
{\cal U}^{\rm disp}_{\pi\pi}(s_a,s_b,s_c)
&=& {1\over 3} \tilde\Delta_{0, \pi\pi}(s_a)
                          \left[3 T(s_a)+T(s_b)+T(s_c) 	\right]\nonumber\\
& & +{1\over 2} \tilde\Delta_{1, \pi\pi}(s_b) \left[T(s_a)-T(s_c) \right]
    +{1\over 2} \tilde\Delta_{1, \pi\pi}(s_c)
\left[T(s_a)-T(s_b)\right]\nonumber\\
& & +{1\over 2} \tilde\Delta_{2, \pi\pi}(s_b) \left[T(s_a)+T(s_c) \right]
    +{1\over 2} \tilde\Delta_{2, \pi\pi}(s_c)
\left[T(s_a)+T(s_b)\right]\nonumber\\
& & -{1\over 3} \tilde\Delta_{2, \pi\pi}(s_a) \left[s_b(t)+T(s_c) \right]
\label{Upipidisp}
\end{eqnarray}
and
\begin{equation}
\label{zweimal}
\tilde\Delta_{I, \pi\pi}(s)= {s^2\over \pi} \int_{4 m_\pi^2}^\infty
ds' {\delta^{I, {\rm ChPT}}(s')\over s'^2(s'-s-i\epsilon)} .
\end{equation}
The phaseshifts $\delta^{I, {\rm ChPT}}$ are the expressions to leading order
in ChPT given in Ref. \cite{GL85c}. $U_{\pi\pi}^{\rm disp}$ contains the
singularities due to $\pi\pi$ intermediate states.
The remainder, $U^{\rm rem}$,
is polynomial except for singularities
due to $\pi\eta$, $K\bar K$ or $\eta\eta$
intermediate states. More explicitly, it is given by
\begin{eqnarray}
{\cal U}^{\rm rem}(s_a,s_b,s_c)
&=&  {1\over 3} \bar\Delta_0(s_a)\left[3 T(s_a)+T(t)+T(s_c) 	\right]\nonumber\\
& & +{1\over 2} \bar\Delta_1(s_b) \left[T(s_a)-T(s_c) 	\right]
    +{1\over 2} \bar\Delta_1(s_c) 	 \left[T(s_a)-T(s_b)\right]
\nonumber\\
& & +{1\over 2} \bar\Delta_2(s_b) \left[T(s_a)+T(s_c) 	\right]
    +{1\over 2} \bar\Delta_2(s_c) 	 \left[T(s_a)+T(s_b)\right]
\nonumber\\
& & +\bar\Delta_3(s_a)
\end{eqnarray}
with
\begin{eqnarray}
\bar\Delta_0(s)&=& {1\over F_\pi^2}\left\{ m_\pi^2 k_{\pi\pi}-
                     (2 k_{\pi\pi}-{1\over 192 \pi^2}) s \right\}\nonumber\\
\bar\Delta_1(s)&=& {1\over F_\pi^2}\left\{ -{1\over 3} k_{\pi\pi}+
                     s m^r_{K K}(s) \right\} \nonumber\\
\bar\Delta_2(s)&=& {1\over F_\pi^2}\left\{ -2 m_\pi^2 k_{\pi\pi}+
                     (k_{\pi\pi}+{1\over 96 \pi^2}) s\right\} \nonumber\\
& &+{(3 s-4 m_K^2)\over 4 F_\pi^2} J^r_{KK}(s)+
                     {m_\pi^2\over 3 F_\pi^2} J^r_{\pi\eta}(s)\nonumber\\
\bar\Delta_3(s_a)&=& {1\over F_\pi^2}\left\{ -{1\over 3}(T(s_b)+T(s_c) 	)
\left[-2 m_\pi^2 k_{\pi\pi}+(k_{\pi\pi}+{1\over 96 \pi^2}) s_a \right]
\right\}\nonumber\\
& &-{s_a(3s_a-4m_\pi^2)\over 4 F_\pi^2(m_\eta^2-m_\pi^2)} J^r_{K K}(s_a)
+{m_\pi^2 (3s_a-4m_\pi^2)\over 3 F_\pi^2 (m_\eta^2-M\pi^2)} J^r_{\pi\eta}(s_a)
\nonumber\\
& &-{m_\pi^2\over 2 F_\pi^2} J^r_{\eta\eta}(s_a)
-{3 s_a (3s_a-4m_K^2)\over 8 F_\pi^2 (s_a-4 m_K^2)} \left[\bar J_{K K}(s_a)
-{1\over 8 \pi^2}\right] .
\end{eqnarray}
The functions $k_{PP}$, $J^r_{P Q}$, $\bar J_{P Q}$ and $M^r_{P Q}$ are
defined in Ref. \cite{GL85a} and are not displayed here explicitely;
when not given, their arguments are as on the left hand sides of the
equations.
${\cal U}_{\pi\pi}^{\rm disp}(s_a,s_b,s_c)$ will be reproduced by
iterating the Khuri-Treiman
equations as discussed before and shown in more detail below. In fact, it is
uniquely determined by unitarity. The rest of the amplitude,
\begin{eqnarray}
\label{resto}
\bar A^{\rm rem}(s_a,s_b,s_c) 	&=&\bar A(s_a,s_b,s_c) 	-
{\cal U}_{\pi\pi}^{\rm disp}(s_a,s_b,s_c) 	\nonumber\\
&=&T(s_a)+{\cal V}(s_a)+{\cal W}(s_a,s_b,s_c)
+{\cal U}^{\rm rem}(s_a,s_b,s_c)
\end{eqnarray}
is smooth in the physical region and it is expanded in a second
order polynomial in $s_a,s_b,s_c$
around the center of the Dalitz plot,
$s_a=s_b=s_c= s_0={1 \over 3}(m_\eta^2+3 m_\pi^2)$:
\begin{equation}
\label{polom}
\bar A{\rm rem}(s_a,s_b,s_c)
=\bar\alpha+\bar\beta (s_a-s_0)+\bar\gamma (s_a-s_0)^2
+\bar\delta (s_b-s_c)^2.
\end{equation}
This form is the most general second order polynomial if the constraint
$s_a+s_b+s_c=3s_0={1 \over 3}(m_\eta^2+3 m_\pi^2)$ and bose
symmetry are
satisfied.
The expansion of $\cal U^{\rm rem}$ is tedious and its analytic form
will not be displayed here.
$T(s_a), {\cal V}(s_a)$ and ${\cal W}(s_a,s_b,s_c)$
are already polynomial and their
contribution to $\bar\alpha$, ..., $\bar\delta$ is
\begin{eqnarray}
\bar\alpha&=&T(s_0)+{\cal V}(s_0)+{\cal W}(s_0,s_0,s_0) \nonumber\\
\bar\beta&=&{3\over (m_\eta^2-m_\pi^2)}\left\{1+a_1+ \right. 3
a_2(m_\eta^2-m_\pi^2)+
a_3(9 m_\eta^2-m_\pi^2) \nonumber\\
& &+{2\over 3}(d_1+{4 m_\pi^2\over(m_\eta^2-m_\pi^2)} \left. d_2)
\right\}-{12 s_0\over F_\pi^2(m_\eta^2-m_\pi^2)} L_3 \nonumber\\
\bar\gamma&=& {6 \over F_\pi^2(m_\eta^2-m_\pi^2)} L_3 \nonumber\\
\bar\delta&=&-{2\over F_\pi^2(m_\eta^2-m_\pi^2)} L_3 .
\label{L3coeff}
\end{eqnarray}
The constants $a_1$, $a_2$, $a_3$, $d_1$ and $d_2$ are tabulated in
Ref. \cite{GL85c}. We note here that the coefficients $\bar\beta$,
$\bar\gamma$ and $\bar\delta$ depend on the low energy constant $L_3$ which
is phenomenologically not  known very accurately
\footnote{Note that $L_3$  is scale independent.}.
Its value extracted from $K_{l 4}$ decays is \cite{Retal89}
\begin{equation}
L_3=(-3.62\pm1.31) \cdot 10^{-3} .
\end{equation}
The error bar on $L_3$ leads to corresponding error bars on $\bar\beta$,
$\bar\gamma$ and $\bar\delta$:
\begin{eqnarray}
\Delta_{|_{L_3}} \bar\alpha &=& 0 \nonumber\\
\Delta_{|_{L_3}} \bar\beta &=& -{12 s_0 \over F_\pi^2 (m_\eta^2-m_\pi^2)}
(\Delta L_3)\approx \mp 0.76 {\rm GeV}^{-2} \nonumber\\
\Delta_{|_{L_3}} \bar\gamma &=& {6 \over F_\pi^2 (m_\eta^2-m_\pi^2)}
(\Delta L_3)\approx \pm 3.18 {\rm GeV}^{-4} \nonumber\\
\Delta_{|_{L_3}} \bar\delta &=& -{2  \over F_\pi^2 (m_\eta^2-m_\pi^2)}
(\Delta L_3)\approx \mp 1.06 {\rm GeV}^{-4}
\end{eqnarray}
We shall comment on the importance of this uncertainty in the determination of
$L_3$ in section 6 where we discuss the phenomenological implications of
our results.

The information needed to fix the subtraction constants of the generalized
Khuri-Treiman equations is contained in coefficients $\bar\alpha$, $\ldots$,
$\bar\delta$, eq. (\ref{L3coeff}), as well as in the corresponding expressions
coming from the expansion of ${\cal U}^{\rm rem}$.

\subsection{Fixing the subtraction constants}

We consider the generalized Khuri-Treiman equations
in the three times subtracted form
given in eqs. (\ref{3.39})-(\ref{3.42})
\footnote{Note the functions $U,V,W$ are not identical
to the functions $\cal U$, etc. introduced above.}.
The $s_1$, $s_2$,... $w_2$ are finite subtraction points,
$f_l^I=\sin \delta_l^I
e^{i\delta_l^I}$ is the partial wave for $\pi\pi$ scattering with angular
momentum $l$ in the isospin channel $I$ and the precise definition of
$\tilde F$ and $\bar F$ has been given in section 3. $P$
is a polynomial of second order in $s_a,s_b,s_c$ with the same
form as eq. (\ref{polom}):
\begin{equation}
P(s_a,s_b,s_c) 	=\alpha+\beta s_a+\gamma s_a^2+\delta (s_b-s_c)^2.
\label{PolynomGL}
\end{equation}
The reason for using three subtractions will become clear soon. Note
that the polynomial $P$ is not equal to $\bar A^{\rm rem}$ because the latter
refers to a twice subtracted dispersion relation (see eq. (\ref{zweimal}))
while $P$ corresponds
to three subtractions.

Now we iterate eq. (\ref{3.40}) by
inserting  on the right hand side latexthe current algebra expression
$F(s_a,s_b,s_c)=T(s_a)$. Moreover, we replace $f_l^I$
by its low energy expansion
\begin{equation}
f_l^I(s_a)\approx \delta_l^{I, {\rm ChPT}}(s_a).
\end{equation}
We observe that this reproduces the absorptive parts of the ChPT
one-loop amplitude due to $\pi\pi$ intermediate states. For isospin $I=0,2$
this is trivial because $T(s_a)$ is linear in $s_a$ and hence
$T(s_b)+T(s_c)$ is again
a first order polynomial in $s_a$. For the $I=1$ $p$-wave $\pi\pi$
intermediate
state the absorptive part is in general given by
\begin{equation}
{\rm Im} F(s_a,s_b,s_c)_{|_{\delta_1^1}}=e^{i\delta_1^1(s_b)}\sin \delta_1^1
(s_b) {3(s_c-s_a)\over 2 K(s_b)} F_1^{s_a}(s_b)+(t\leftrightarrow s_c)
\end{equation}
where $K(s_b)$ is the Kacser function and $F_1^{s_a}(s_b)$ is the p-wave
projection
in the $s_b$-channel of the function $F(s_a,s_b,s_c)$ defined in section 3.
Setting
$T(s_a)=a+b s_a$, the p-wave projection in the $s_b$-channel is
calculated to be
\begin{equation}
T_1^{s_a}(s_b)=-{1\over 3} K(s_b) b.
\label{Tppro}
\end{equation}
Thus the first iteration with the approximations given above yields
\begin{equation}
{\rm Im} F(s_a,s_b,s_c)_{|_{\delta_1^1}}=\delta_1^{1, {\rm ChPT}} {1\over 2}
(T(s_a)-T(s_c) 	)+(s_b\leftrightarrow s_c) 	.
\end{equation}
This is indeed the result of the one-loop
calculation obtained in Ref. \cite{GL85c}.

Turning now to the dispersive part, we shall show how the real part of
$\cal U_{\pi\pi}^{\rm disp}$, eq. (\ref{Upipidisp}),
is reproduced in the first
iteration of the generalized Khuri-Treiman equations. In order to make the
argument more transparent, let us consider first the simpler case where all
subtractions are taken at zero. The term ${U}(s_a)$ on the right
hand side of eq. (\ref{3.39}) then reads, after the first iteration
\begin{equation}
{\rm Re}\, U^{\rm iter}(s_a)={s^3\over \pi} P\int_{4 m_\pi^2}^\infty ds'
{\delta_0^{0, {\rm ChPT}}(s')(\hat a+\hat b s')-\delta_0^{2, {\rm ChPT}}
(\bar a+\bar b s') \over s'^3(s'-s_a-i\epsilon)}
\label{ReUiter}
\end{equation}
with
\begin{eqnarray}
\label{ahut}
\hat a &=& {5\over 3}a+b s_0,\qquad\qquad \hat b = {2\over 3} b \nonumber\\
\bar a &=& {2\over 3}a+b s_0,\qquad\qquad \bar b =-{1\over 3} b
\end{eqnarray}
and
\begin{equation}
\label{anorm}
a=1-{3 s_0\over m_\eta^2-m_\pi^2}, \qquad\qquad b={3\over m_\eta^2-m_\pi^2}.
\end{equation}
In eq. (\ref{ReUiter}), $P$ denotes the principal value of the integral;
the combinations in eq. (\ref{ahut}) simply stand for the appropriate
isospin projections.
Using the decomposition
\begin{equation}
\label{drzw}
{s^3\over \pi} P\int ds'{\delta^I(s')(s'-z)\over s'^3(s'-s-i\epsilon)}=
(s-z){s^2\over \pi} P\int ds'{\delta^I(s')\over s'^2(s'-s-i\epsilon)}
+z{s^2\over \pi} P\int ds'{\delta^I(s')\over s'^3}
\end{equation}
eq. (\ref{ReUiter}) can be brought into the form
\begin{eqnarray}
{\rm Re}\, {U}^{\rm iter}(s_a)&=&\left[ T(s_a)+{1\over 3}(T(s_b)+T(s_c))\right]
{\rm Re}\tilde\Delta_{0, \pi\pi}(s_a) \nonumber\\
& & -\hat a {s_a^2\over\pi} P\int ds'{\delta_0^{0, {\rm ChPT}}(s')\over s'^3}
\nonumber\\
& & -{1\over 3}\left[T(s_a)+T(s_c) \right] {\rm Re}\tilde\Delta_{2,
\pi\pi}(s_a)
\nonumber\\
& & +\bar a {s^2\over\pi} P\int ds'{\delta_0^{2, {\rm ChPT}}(s')\over s'^3}.
\label{Discre1}
\end{eqnarray}
We thus have transformed ${\rm Re}\, U^{\rm iter}(s)$ into the first
and last term of $\cal U_{\pi\pi}^{\rm disp}$ plus two polynomial terms.
In other words, eq. (\ref{drzw}) serves to transform a three times
subtracted relation into a twice subtracted one. Likewise,
we may decompose
\begin{eqnarray}
{\rm Re}\left(V(s_b)+V(s_c) 	\right)^{\rm iter}
&=&{1\over 2}\left[ T(s_a)+T(s_c) 	\right]
{\rm Re}\tilde\Delta_{2, \pi\pi}(s_b) \nonumber\\
& & -\tilde a {s_b^2\over\pi} P\int ds'{\delta_0^{2, {\rm ChPT}}(s')\over s'^3}
+(s_b \leftrightarrow s_c)
\label{Discre2}
\end{eqnarray}
with
\begin{equation}
\tilde a = a+{3\over 2}b s_0.
\end{equation}
Finally, using (\ref{Tppro}) and the definition of $\hat F(s')$,
eq. (\ref{3.37}), we see
that the $I=1$ contribution in the generalized Khuri-Treiman equations
yields precisely the terms in proportion to $\tilde \Delta_{1, \pi\pi}$
in eq. (\ref{Upipidisp}):
\begin{equation}
{\rm Re}\left((s_a-s_c){W}(s_b)+(s_a-s_b){W}(s_c)\right)^{\rm iter}
={1\over 2}\left[ T(s_a)-T(s_c) 	\right]
{\rm Re}\tilde\Delta_{1, \pi\pi}(s_b) +(s_b \leftrightarrow s_c) 	.
\end{equation}

We see that linking the dispersion technique
to the one-loop ChPT calculation allows to subtract $U$,$V$ and
$W$ separately avoiding the problem (mentioned previously)
that these points may lie outside
the physical region.

We are now ready to determine the subtraction polynomial ${P}$ of the
triple subtracted Khuri-Treiman equations.  After
the first iteration, the amplitude can be written as
\begin{equation}
F(s_a,s_b,s_c)^{\rm iter}={P}(s_a,s_b,s_c)+{\cal U}_{\pi\pi}
^{\rm disp}(s_a,s_b,s_c)
+{R}(s_a,s_b,s_c) 	.
\end{equation}
If all subtractions are taken at zero, ${R}$ is given by eqs.
(\ref{Discre1})- (\ref{Discre2}), i.e.
\begin{eqnarray}
{R}(s_a,s_b,s_c) 	&=&
-\hat a{s_a^2\over\pi} P\int ds'{\delta_0^{0, {\rm ChPT}}(s')\over s'^3}
\nonumber\\
& & +\left( \bar a {s_a^2\over \pi}-\tilde a {s_b^2+s_c^2\over\pi} \right)
P\int ds'{\delta_0^{2, {\rm ChPT}}(s')\over s'^3}.
\end{eqnarray}
Requiring the matching of the first iteration of the generalized
Khuri-Treiman
equations with the one-loop amplitude from ChPT we obtain
\begin{eqnarray}
\label{Pfix}
{P}(s_a,s_b,s_c) 	&=&\left\{\bar A(s_a,s_b,s_c)
-{\cal U}_{\pi\pi}^{\rm disp}-R(s_a,s_b,s_c)
\right\}_{\rm expand} \\
&=& \bar\alpha+\bar\beta (s_a-s_0)+\bar\gamma (s_a-s_0)^2+\bar\delta
(s_b-s_c)^2-{R}(s_a,s_b,s_c) \nonumber
\end{eqnarray}
where the subscript ``expand'' means a Taylor expansion up to second order
in $s_a,s_b,s_c$; in fact the expression in the curly brackets
is just $\bar A^{\rm rem}-R$.

Eq. (\ref{Pfix}) and its generalization for finite subtractions points
described below is the main result of this section. It describes our method
of using the ChPT one-loop amplitude (see however below)
to fix the subtraction constants of the
generalized Khuri-Treiman equations. In order to make the
procedure transparent,
that is show explicitely how the higher corrections unitarize the amplitude,
we were forced to subtract the Khuri-Treiman equations three times. Otherwise
the integrals  occuring in ${R}(s_a,s_b,s_c)$ would not converge.
By construction, the first step in an iteration of the Khuri-Treiman
equations reproduces the ChPT one-loop result in the physical decay region
to very good accuracy. The numerical solution of the generalized
Khuri-Treiman equations as attempted in section 5 can therefore
be interpreted
as a correction on top of the ChPT one-loop amplitude due to all possible
two-body final state interactions of pions.

There remains the important question of higher order corrections to
the one-loop results which may shift the subtraction constants
substantially. We will discuss adress this issue in section 6. Here
we only note that also in this case the above framework of three subtractions
can be employed if the necessary modifications are made.

\subsection{Finite subtraction points}

In the preceeding subsection all three subtractions have been taken at zero
for the sake of simplicity. Here we shall describe the modifications which
arise if the subtractions are taken at finite values.
We do this for two reasons: Since the dispersion relations do not
fix the subtractions points, varying them over a certain domain
gives an estimate of the error of our
procedure (see also the discussion in section 6).
Second,
it is numerically favorable to have subtraction
points which do not coincide
when we solve the Khuri-Treiman equations iteratively.


Introducing now finite subtraction points and
performing essentially the same steps as before, we find that only the
function $R(s_a,s_b,s_c) $ is modified. It explicitly depends on
subtraction points $s_i, v_i$, i=1,2,3 and $w_i$, i=1,2 and is given as
\begin{eqnarray}
R(s_a,s_b,s_c; 	s_i)&=& \hat b R^{(0)}(s_a;s_1,s_2,s_3;
-{\hat a\over\hat b}) -\bar b R^{(2)}(s_a;v_1,v_2,v_3;
-{\bar a\over\bar b}) \nonumber\\
& & +\tilde b R^{(2)}(s_b;v_1,v_2,v_3;
-{\tilde a\over\tilde b}) +\tilde b R^{(2)}(s_c;v_1,v_2,v_3;
-{\tilde a\over\tilde b}) \nonumber\\
& & +b R^{(1)}(s_a,s_b,s_c;w_1,w_2).
\label{Rfinite}
\end{eqnarray}
The explicit form of functions $R^{(I)}$, $I=0,1,2$ is given in Appendix D.

The subtraction polynomial is then again given by eq. (\ref{Pfix}) with
$R$ taken from (\ref{Rfinite}). It depends explicitly on the subtraction
points $s_i$, $v_i$ and $w_i$. However, by construction this dependence on
the subtraction points is
counterbalanced by a corresponding dependence of the dispersive part of the
generalized Khuri-Treiman equations, provided only the first step of the
iteration
described above is performed. The procedure of fixing subtraction constants is
in this sense independent on the choice of the subtraction points. Up to the
first iteration, the ChPT one-loop amplitude is reproduced, in the physical
region, for any value $s_1$, ..., $w_2$ in the low energy regime. Beyond the
first iteration this is no longer true.

Finally we give numerical values for coefficients $\alpha$, ..., $\delta$ of
the subtraction polynomial $P$. We consider the sets of subtraction points
$SP={s_1=v_1=w_1, s_2=v_2=w_2, s_3=v_3}$ displayed in Table 1. The input
parameters we
use are $m_\pi=140 {\rm MeV}$, $m_\eta=549 {\rm MeV}$, $F_\pi=92.4 {\rm MeV}$,
$F_K=114 {\rm MeV}$, $L_3=(-3.62\pm 1.31) 10^{-3}$.
The calculated coefficients
$\alpha$, ..., $\delta$ for these sets of subtraction points are given in
Table 2.

The dependence on the choice of subtraction points is substantial
and we shall
discuss how it propagates into the final numerical solution
in section 6. Note that error bars due to the uncertainty in the
determination of $L_3$ are correlated. For instance, the value of
$P(s_a,s_b,s_c) $
at the center of the Dalitz plot does not depend on $L_3$ -- the
errors for $\alpha$, ..., $\delta$ cancel for this quantity. Also it appears
that for subtraction points $SP\approx 0$,
$\gamma$ is rather small. In the physical region, the term in proportion
to $\gamma$ contributes only $\approx$ 2\% to the subtraction polynomial.
The error bar on $\gamma$ due to the uncertainty in $L_3$ is therefore
phenomenologically less important.

\subsection{Comments}
We have swept over several subtle points rather briskly and would like
to come back to them.

\begin{enumerate}
\item
The polynomial $P$ depends on four constants.
On the other hand, we have subtracted the three functions $U$, $V$
and $W$ separately; this implies $3+3+2=8$ constants. It is clear
from the form of $P$ that there are only four physically relevant
parameters and that there are redundant parameters. The reason for
this is found in a general invariance of the dispersion
relations: they fix a function only up to
a polynomial; therefore, we can 'shift' certain constants
from one function to the other. For instance, we can redefine
in this way two constants in $W$ and two in $V$ (the constant and the
linear terms). Thus, only four constants remain.

\item
The asymptotic behaviour of the amplitude has been discussed only briefly.
Taking three subtractions we want to make sure that the actual behaviour
in the asymptotic region does not matter for the solution in the physical
decay region. There is however an other subtlety which we should mention.
Mathematically speaking, the solution to the integral equations
(\ref{3.39}-\ref{3.43}) is
not completely determined by specifying the asymptotic behaviour of the
amplitude. The number of parameters (extracted from ChPT) to be input depends
also on the asymptotic behaviour of the $\pi\pi$-phase shift.
\footnote{We are indebted to H. Leutwyler for pointing out this to us. A
more complete discussion of these issues can be found in Ref. \cite{Heiri}.}
Let us consider the twice subtracted relation for definiteness and
omit $V$ and $W$. The solution to this problem can be written in the form
\begin{equation}
F(s)=P(s)+\Ph_{10}(s)+\psi_0(s)
\end{equation}
where $\Ph_{10}$ and $\psi_0$ are defined in eqs. (\ref{5.8}-\ref{5.13}).
Now, if the phase goes to zero for large $s$, the Omnes factor is
constant; correspondingly, the amplitude rises linearly with $s$.
On the other hand, if the phase approaches $\pi$, the Omnes factor
decays like $s^{-1}$ and the amplitude tends towards a constant. Stated
differently, a given asymptotic behaviour of the amplitude, say $\sim$ const.,
requires one or two subtractions for the two limiting cases of the phase shift
respectively. This ambiguity in the phase is reflected in an additional term
\begin{equation}
{Q(s)\over D(s)} (s-s_1)(s-s_2)(s-s_3), \qquad Q(s)\ {\rm polynomial},
\label{Qterm}
\end{equation}
which can be added to the approximate solution $\Ph_{10}$, c.f. eq.
(\ref{5.28}). It introduces an uncertainty in the final solution which
should be well controlled since the assumptions entering the analysis are
not rigorous. We have estimated this uncertainty by modelling the phase in
the high energy region. The details as well as the resulting error bars
on our final answer are given in section 6.
\end{enumerate}

\Section{The iteration}

With reasonable assumptions we now build up an iterative scheme
for the numerical solution of the set of coupled equations
for the
 projected amplitudes $U,V,W$. It consists of two distinct
kinds of iterations. The first one accumulates contributions to
$V$ and $W$ for a corresponding part of $U$ while the second
yields the contributions to $U$ itself.

The first type of iteration relies on the assumption that
to lowest order only $U$ contributes, but not $V$ and $W$.
It is motivated by the fact that
 pion-pion scattering at low enegies is dominated by the $s$-wave, $I=0$
channel as the $s$-wave, $I=2$ contribution is suppressed
by a much smaller phase shift
$f^{2}_{0}\ll f^{0}_{0}$. The $I=1$ contribution comes
from a $p$-wave scattering and therefore is also suppressed.
We may thus expect that
in comparison to $V$ and $W$, $U$
 will still yield the dominant contribution
 to $F$ in our refined analysis.

To organize this first iteration we take the terms with the
phase shifts $f^{2}_{0},f^{1}_{1}$ as perturbations and
introduce accordingly a counting parameter $\la$.
$f^{0}_{0}$ is taken to be of $O(\la^{0})$ whereas
\beqq f^{2}_{0}&\ar&\la f^{2}_{0} \\
f^{1}_{1}&\ar&\la f^{1}_{1} \nonumber \eeqq
are treated as small parameters. In principle,
one could introduce different counting parameters
for the two isospin channels, but for sufficently
many iterations in $\la$, this is irrelevant.
Expansion of $U,V$ and $W$ in a series
in $\la$ yields
\beqq U(s_{a})&=&\sum_{0}^{\infty}\la^{k}U_{k}(s_{a}) \nonumber \\
V(s_{a})&=&\sum_{0}^{\infty}\la^{k}V_{k}(s_{a}) \\
W(s_{a})&=&\sum_{0}^{\infty}\la^{k}W_{k}(s_{a}) \nonumber \eeqq
where we set by assumption $V_{0}(s_{a})=W_{0}(s_{a})=0$.
We insert these expansions in the integral equations
eqs. (\ref{3.47}), (\ref{3.48}) and (\ref{3.49}) and
obtain by equating equal powers of $\la$
\beqq \label{5.3}
U_{k}(s_{a})&=&\frac{\prod_{i=1}^{3}(s_{a}-s_{i})}{\pi}
\intgr\frac{ds'_{a}}{\prod_{j=1}^{3}(s'_{a}-s_{j})(s'_{a}-s_{a}-i\ep)}
\nonumber \\
& & \quad\cdot\{f^{0*}_{0}(s'_{a})\,[\{{\overline P}(s'_{a})+\frac{2}{3}
{\tilde P}(s'_{a})\}\de_{k0}+U_{k}(s'_{a})+\frac{2}{3}
{\overline U}_{k}(s'_{a}) \nonumber \\
& & \quad\quad+\frac{2}{3}V_{k}(s'_{a})+\frac{8}{3}
{\overline V}_{k}(s'_{a})
+\frac{4}{3}{\overline{(s_{c}-s_{a})W_{k}}}\,(s'_{a})] \\
& & \quad-f^{2*}_{0}(s'_{a})\,[\frac{2}{3}
{\tilde P}(s'_{a})\de_{k1}+\frac{2}{3}
{\overline U}_{k-1}(s'_{a})+\frac{2}{3}V_{k-1}(s'_{a}) \nonumber \\
& & \quad\quad+\frac{2}{3}{\overline V}_{k-1}(s'_{a})
+\frac{2}{3}{\overline{(s_{a}-s_{c})W_{k-1}}}\,(s'_{a})]\},
 \nonumber \eeqq
\beqq \label{5.4}
V_{k}(s_{a})&=&\frac{\prod_{i=1}^{3}(s_{a}-s_{i})}{\pi}
\intgr\frac{ds'_{a}}{\prod_{j=1}^{3}(s'_{a}-s_{j})(s'_{a}-s_{a}-i\ep)}
\nonumber \\
& &  \quad\cdot f^{2*}_{0}(s'_{a})
\,[{\tilde P}(s'_{a})\de_{k1}+
{\overline U}_{k-1}(s'_{a})+V_{k-1}(s'_{a}) \\
& & \quad\quad+{\overline V}_{k-1}(s'_{a})+
{\overline{(s_{a}-s_{c})W_{k-1}}}\,(s'_{a})]
\nonumber \eeqq
and
\beqq \label{5.5} W_{k}(s_{a})&=&\frac{(s_{a}-s_{1})(s_{a}-s_{2})}{3\pi}
\intgr\frac{ds'_{a}}{(s'_{a}-s_{1})(s'_{a}-s_{2})(s'_{a}-s_{a}-i\ep)}
\nonumber \\ & &
\quad\cdot f^{1*}_{1}(s'_{a})\,[{\hat P}(s'_{a})\de_{k1} \\ & &  \quad\quad
+\frac{3}{4K^{2}(s'_{a})}\{{\overline{(s_{c}-s_{b})U_{k-1}}}\,(s'_{a})
+{\overline{(s_{b}-s_{c})V_{k-1}}}\,(s'_{a}) \nonumber \\
& &\quad\quad +{\overline{(s_{b}-s_{c})^{2}}}
\cdot W_{k-1}(s'_{a})+{\overline{(s_{b}-s_{c})
(s_{a}-s_{c})W_{k-1}}}\,(s'_{a})\}] \nonumber \\
& &+\mbox{perm.} \nonumber \eeqq
The structure of the iterations becomes clear if
 we rewrite the equation for $U_{k}$ in the form
\beqq \label{5.6} U_{k}(s_{a})&=&\Ph_{0k}(s_{a})+
\frac{\prod_{i=1}^{3}(s_{a}-s_{i})}{\pi}
\intgr\frac{ds'_{a}}{\prod_{j=1}^{3}(s'_{a}-s_{j})(s'_{a}-s_{a}-i\ep)}
\nonumber \\ & & \quad\cdot f^{0*}_{0}(s'_{a})\,[U_{k}(s'_{a})+\frac{2}{3}
{\overline U}_{k}(s'_{a})] \eeqq
introducing the combination
\beqq \label{5.7} \Ph_{0k}(s_{a})&\equiv&
\frac{\prod_{i=1}^{3}(s_{a}-s_{i})}{\pi}
\intgr\frac{ds'_{a}}{\prod_{j=1}^{3}(s'_{a}-s_{j})(s'_{a}-s_{a}-i\ep)}
\nonumber \\ & & \quad\cdot\{f^{0*}_{0}(s'_{a})\,[\{{\overline P}(s'_{a})
+\frac{2}{3}{\tilde P}(s'_{a})\}\de_{k0}+\frac{2}{3}V_{k}(s'_{a}) \nonumber \\
& & \quad\quad+\frac{8}{3}{\overline V}_{k}(s'_{a})
+\frac{4}{3}{\overline{(s_{c}-s_{a})W_{k}}}\,(s'_{a})] \\
& & \quad-f^{2*}_{0}(s'_{a})\,[\frac{2}{3}{\tilde
P}(s'_{a})\de_{k1}+\frac{2}{3}
{\overline U}_{k-1}(s'_{a})+\frac{2}{3}V_{k-1}(s'_{a}) \nonumber \\
& & \quad\quad+\frac{2}{3}{\overline V}_{k-1}(s'_{a})
+\frac{2}{3}{\overline{(s_{a}-s_{c})W_{k-1}}}\,(s'_{a})]\}.
\nonumber \eeqq
For given $U_{k-1},V_{k-1},W_{k-1}$ and their respective
 projections we may use eqs.  (\ref{5.4})-(\ref{5.5}) to
determine $V_{k}$ and $W_{k}$. This allows to compute the
subtraction function $\Ph_{0k}$ as given in eq. (\ref{5.7}).
If we can also solve eq. (\ref{5.3}) for $U_{k}$, the
iteration step is complete and we may perform it once again for
$k$ instead of $k-1$. Note that the lowest contributions
$U_{0},V_{0}$ and $W_{0}$ are completely determined by the
different projections of the subtraction polynomial $P$
containing the input information of ChPT on the decay
amplitude.


As the process converges rapidly, we will have to perform a small number
of iterations, typically four or five, to obtain a precision below
one part in thousand.
Of course we take now $\la=1$ as it was introduced as
a counting parameter only and may sum the different contributions
to obtain the fully iterated $U,V$ and $W$.

We turn to the second kind of iteration which yields the yet
undetermined $U_{k}$ from eq. (\ref{5.3}). We thereby follow
the discussion of Bronzan \cite{BR} and
Neveu and Scherk \cite{NS70} extending it to our case. As we
are dealing with one variable only we omit the usual subscripts here.

If in the equation (\ref{5.6}) for $U_{k}$ the second term
in the square bracket were
absent, we would have to deal with an
Omn$\grave{\mbox{e}}$s type of integral equation
describing ordinary two pion-pion rescattering without
rediffusion terms coming from the third pion in the final
state. In our case making use of solvable Omn$\grave{\mbox{e}}$s
equations we are able to recast the equation for $U_{k}$ in a form
which is now accessible for a second kind of
iteration \cite{BR,NS70}.
To this end we introduce two auxiliary functions, the first of
which is defined by

\beqq \label{5.8}
\Ph_{1k}(s)&\equiv&\Ph_{0k}(s)+\frac{\prod_{i=1}^{3}(s-s_{i})}
{\pi}\intgr\frac{ds'}{\prod_{j=1}^{3}(s'-s_{j})(s'-s-i\ep)}
\nonumber \\ & & \quad\quad\quad\quad\cdot f^{0*}_{0}(s')\,\Ph_{1k}(s).
\eeqq
This equation is indeed solvable in terms of $\Ph_{0k}$ \cite{JDJ}
and yields
\beqq  \label{5.9}
\Ph_{1k}(s)&=&\Ph_{0k}(s)+\frac{\prod_{i=1}^{3}(s-s_{i})}
{\pi D(s_{+})}\intgr\frac{ds'}{\prod_{j=1}^{3}(s'-s_{j})(s'-s-i\ep)}
\nonumber \\ & & \quad\quad\quad\quad\cdot f^{0}_{0}(s')D(s'_{+})\,\Ph_{0k}(s')
\eeqq
where $s_{+}=s+i\ep$ and $D(s)$ is just the Omn$\grave{\mbox{e}}$s
function corresponding to the phase shift $\de^{0}_{0}$
\beq \label{5.10} D(z)= e^{-\frac{1}{\pi}\intgr
ds'\frac{\de^{0}_{0}(s')}{s'-z}}.
\eeq
The second auxilary function $\psi_{k}$ is then defined by
\beq \label{5.11} U_{k}(s)+\frac{2}{3}
{\overline U}_{k}(s)\equiv\Ph_{1k}(s)+\psi_{k}(s).
\eeq
Using this definition twice and
$\Ph_{1k}$ in eq. (\ref{5.8}), we obtain an equation for $\psi_{k}$
\beqq
\psi_{k}(s)&=&\frac{2}{3}{\overline U}_{k}(s)+
\frac{\prod_{i=1}^{3}(s-s_{i})}{\pi}
\intgr\frac{ds'}{\prod_{j=1}^{3}(s'-s_{j})(s'-s-i\ep)}
\nonumber \\ & & \quad\quad\quad\quad\cdot f^{0*}_{0}(s')\,\psi_{k}(s)
\eeqq
which is again of the Omn$\grave{\mbox{e}}$s type and has
a solution in terms of the unknown ${\overline U}_{k}$
\beqq \label{5.13}
\psi_{k}(s)&=&\frac{2}{3}{\overline U}_{k}(s)+
\frac{\prod_{i=1}^{3}(s-s_{i})}{\pi D(s_{+})}
\intgr\frac{ds'}{\prod_{j=1}^{3}(s'-s_{j})(s'-s-i\ep)}
\nonumber \\
& & \quad\quad\quad\quad\cdot f^{0}_{0}(s')D(s'_{+})\,
\frac{2}{3}{\overline U}_{k}(s'). \eeqq
Eqn.(\ref{5.11}) is next rewritten as
\beq U_{k}(s)=\Ph_{1k}(s)+\psi_{k}(s)-\frac{2}{3}
{\overline U}_{k}(s). \eeq
But the difference of the last two terms is just
 given by the integral on the r.h.s. of eq. (\ref{5.13})
 so that we finally obtain the desired form of the equation
for $U_{k}$ \cite{BR,NS70}
\beqq \label{5.15} U_{k}(s)&=&\Ph_{1k}(s)+
\frac{\prod_{i=1}^{3}(s-s_{i})}{\pi D(s_{+})}
\intgr\frac{ds'}{\prod_{j=1}^{3}(s'-s_{j})(s'-s-i\ep)}
\nonumber \\ & & \quad\quad\quad\quad\cdot f^{0}_{0}(s')D(s'_{+})\,
\frac{2}{3}{\overline U}_{k}(s'). \eeqq
$\Ph_{1k}$ is already determined in eq. (\ref{5.9}) as an
Omn$\grave{\mbox{e}}$s inversion. We may now account for
the second term in eq. (\ref{5.15}) by iteration, noting
that at every step one multiplies with the phase shift $f_{0}^{0}$.
Introducing the counting parameter $\mu$
\beq f^{0}_{0}\ar\mu f^{0}_{0} \eeq
we expand $U_{k}$ in a series in $\mu$
\beq
\label{ukuk}
U_{k}(s)=\sum_{1}^{\infty}\mu^{m-1}\Ph_{mk}(s) \eeq
and obtain for $m>1$
\beqq \Ph_{mk}(s)&=&\frac{\prod_{i=1}^{3}(s-s_{i})}
{\pi D(s_{+})}\intgr\frac{ds'}{\prod_{j=1}^{3}(s'-s_{j})(s'-s-i\ep)}
\nonumber \\ & & \quad\quad\quad\quad\cdot f^{0}_{0}(s')D(s'_{+})\,\frac{2}{3}
{\overline\Ph}_{(m-1)k}(s'). \eeqq

Again a small number of iterations is sufficient as
the process here converges rapidly, too.
Taking now $\mu=1$, we sum up the different
contributions to obtain the full iterated $U_{k}$.

Next, we recast the main contribution $\Ph_{1k}$ to $U_{k}$
in a more convenient form. With its use we may give $\Ph_{10}$
explicitly in terms of some projections of the subtraction
polynomial $P$ and the Omn$\grave{\mbox{e}}$s function.

We start with the dispersion representation for $\Ph_{0k}$
as given in eq. (\ref{5.7})
\beq \label{5.19} \Ph_{0k}(s)=\frac{\prod_{i=1}^{3}
(s-s_{i})}{\pi}\intgr\frac{ds''}
{\prod_{j=1}^{3}(s''-s_{j})(s''-s-i\ep)} A(s'') \eeq
where the function $A$ is given by
\beqq A(s)&\equiv&f^{0*}_{0}(s)\,[\{{\overline P}(s)
+\frac{2}{3}{\tilde P}(s)\}\de_{k0}+\frac{2}{3}V_{k}(s) \nonumber \\
& & \quad\quad+\frac{8}{3}{\overline V}_{k}(s)
+\frac{4}{3}{\overline{(s_{c}-s_{a})W_{k}}}\,(s)] \\
& & \quad-f^{2*}_{0}(s)\,[\frac{2}{3}{\tilde P}(s)\de_{k1}+\frac{2}{3}
{\overline U}_{k-1}(s)+\frac{2}{3}V_{k-1}(s) \nonumber \\
& & \quad\quad+\frac{2}{3}{\overline V}_{k-1}(s) +\frac{2}{3}
{\overline{(s_{a}-s_{c})W_{k-1}}}\,(s)] \nonumber \eeqq
and insert it into the dispersion integral for $\Ph_{1k}$ as
displayed in eq. (\ref{5.9}). First, we focus only on the second
term  which becomes
\beqq \label{5.21}
& &\frac{\prod_{i=1}^{3}(s-s_{i})}{\pi D(s_{+})}
\intgr ds''\frac{A(s'')}{\prod_{j=1}^{3}(s''-s_{j})}
\nonumber \\ & & \quad\quad \cdot\frac{1}{\pi}
\intgr ds'\frac{f^{0}_{0}(s')D(s'_{+})}{(s''-s'-i\ep)(s'-s-i\ep)}.
\eeqq
As the discontinuity of the  Omn$\grave{\mbox{e}}$s function
across its cut is given by
\beq \label{5.22} D(s_{+})-D(s_{-})=-2iD(s_{+})f^{0}_{0}(s)
=-2iD(s_{-})f^{0*}_{0}(s) \eeq
we may represent it as
\beq D(z)=-\frac{1}{\pi}
\intgr ds'\frac{D(s'_{+})f^{0}_{0}(s')}{s'-z}.
\eeq
With this representation, the $s'$-integration in
eq. (\ref{5.21}) may be performed and yields the difference
\beq \frac{1}{s''-s-i\ep}[D(s''_{-})-D(s_{+})]
\eeq
so that the expression (\ref{5.21}) becomes now
\beq-\Ph_{0k}(s)+\frac{\prod_{i=1}^{3}(s-s_{i})}{\pi D(s_{+})}
\intgr ds''\frac{D(s''_{-})A(s'')}{\prod_{j=1}^{3}(s''-s_{j})(s''-s-i\ep)}.
\eeq
Inserting this result in eq. (\ref{5.19}) $\Ph_{0k}$ cancels and one obtains
\beqq \label{5.26} \Ph_{1k}(s)&=&\frac{\prod_{i=1}^{3}(s-s_{i})}{\pi D(s_{+})}
\intgr\frac{ds'}{\prod_{j=1}^{3}(s'-s_{j})(s'-s-i\ep)}
\nonumber \\ & & \cdot D(s'_{-})\{f^{0*}_{0}(s')\,[\{{\overline P}(s')+
\frac{2}{3}{\tilde P}(s')\}\de_{k0}+\frac{2}{3}V_{k}(s') \nonumber \\
& & \quad\quad+\frac{8}{3}{\overline V}_{k}(s')
+\frac{4}{3}{\overline{(s_{c}-s_{a})W_{k}}}\,(s')] \\
& & \quad-f^{2*}_{0}(s')\,[\frac{2}{3}{\tilde P}(s')\de_{k1}+
\frac{2}{3}
{\overline U}_{k-1}(s')+\frac{2}{3}V_{k-1}(s') \nonumber \\
& & \quad\quad+\frac{2}{3}{\overline V}_{k-1}(s') +
\frac{2}{3}{\overline{(s_{a}-s_{c})W_{k-1}}}\,(s')]\}. \nonumber
\eeqq

Up to now we only relied on the dispersion representation of
$\Ph_{0k}$ such that the discussion is valid for general $k$.
 From now on it is assumed that $D(z)$ is the only function
with a cut
under the integral (\ref{5.26}). This is correct
only for $k=0$. Using eq. (\ref{5.22}) to rewrite $f^{0*}_{0}(s)
D(s_{-})$ we may recast the dispersion integral along the cut
$L$ as a contour integral along $C$ starting at $\infty$ and
going down to $4$ lying infinitesimally below the cut $L$,
turning there and going back to $\infty$ infinitesimally
above the real axis such that
\beqq
\label{neveu}
\Ph_{10}(s)&=&-\frac{\prod_{i=1}^{3}(s-s_{i})}{2i\pi D(s_{+})}
\int_{C}\frac{dz'}{\prod_{j=1}^{3}(z'-s_{j})(z'-s-i\ep)}
\nonumber \\
& & \quad\quad\quad\cdot D(z')\,[{\overline P}(z')+
\frac{2}{3}{\tilde P}(z')] \\
& & +\prod_{i=1}^{3}(s-s_{i}) {Q(s)\over D(s)}. \nonumber
\eeqq
$Q(s)$ is a polynomial which is not restricted as long as
no asymptotic boundary conditions are imposed.
The form (\ref{neveu}) is easily evaluated with the help of the
residue calculus and yields after a little algebra the final result
\beqq
\label{5.28}
& &\Ph_{10}(s)=[{\overline P}(s)+\frac{2}{3}{\tilde P}(s)]
\cdot\frac{1-D(s)}{D(s)}
\nonumber \\
& &\quad+[{\overline P}(s_{1})+\frac{2}{3}{\tilde P}(s_{1})]\cdot
\frac{(s-s_{2})(s-s_{3})}{(s_{1}-s_{2})(s_{1}-s_{3})}
\cdot\frac{D(s_{1})-1}{D(s)}
\nonumber \\
& &\quad+[{\overline P}(s_{2})+\frac{2}{3}{\tilde P}(s_{2})]\cdot
\frac{(s-s_{1})(s-s_{3})}{(s_{2}-s_{1})(s_{2}-s_{3})}\cdot
\frac{D(s_{2})-1}{D(s)} \\
& &\quad+[{\overline P}(s_{3})+\frac{2}{3}
{\tilde P}(s_{3})]\cdot
\frac{(s-s_{1})(s-s_{2})}{(s_{3}-s_{1})(s_{3}-s_{2})}
\cdot\frac{D(s_{3})-1}{D(s)}
\nonumber \\
& & +\prod_{i=1}^{3}(s-s_{i}) {Q(s)\over D(s)}. \nonumber
\eeqq
Note that it is of course possible to set two or more
of the subtraction points $s_{i}$ in $\Ph_{10}$ equal.
As one has then to deal with multiple poles the expression
for the corresponding residue becomes more involved.

We have built up an iterative scheme for the numerical
solution of the set of coupled integral equations for
the projected amplitudes $U,V$ and $W$ and now turn to a
short description of the numerical aspects of the iterations.

To obtain the decay amplitude with the inclusion of the
final state interactions as described by the Khuri-Treiman
equation we need to know the three functions $U(z),V(z),W(z)$
only for physical values of $z$ lying infinitesimally above
the real axis, $z=s+i\ep$. But in both types of iterations
described above, projections of $U,V$ and $W$ have to be computed.
Since the corresponding projection integrals involve contours
lying in the complex plane, the functions must be known in
a certain domain
around the real axis.

As discussed in section 3 there is one $s_{a}$-region,
namely $(m-1)^{2}<s_{a}<(m+1)^{2}$, in which the projection
integral has to be performed along a contour joining two
points lying in the upper and lower complex half planes
respectively which must not intersect the real axis from
$4$ up to $\infty$ as the integrands have a cut there.
As they are analytic elsewhere we are free to choose those
paths which minimize the number of necessary lattice points
in the complex plane. The two possible choices join the
respective end points as given in eq. (\ref{3.35})
along the path the end points themselves describe as $s_{a}$
increases from $(m-1)^{2}$ to $(m+1)^{2}$. In a second
$s_{a}$-region with $\frac{m^{2}-1}{2}<s_{a}<(m-1)^{2}$
we have to perform integrals along paths lying infinitesimally
above and below the real axis such that we have to know all
the functions involved for $z=s-i\ep$ if $4<s<m+1$.

As we are doing numerical computations, we have to replace the
connected domains just discussed by a lattice. The spacing of
its points is adapted to the accuracy required and to the
possible occurence of numerically problematical points.
Of course we will cut off the different integrals at some
large value of the integration variable which is determined
on one hand by the phenomenological knowledge of the pion
scattering phases involved and on the other by demanding
that numerical results do not change sensibly if one
varies the cut-off. The corresponding cut-off for
negative values of $s_{a}$ is then automatically
fixed by eqs. (\ref{3.35}). As a result we have to
know all the functions on a lattice of points
lying in the complex plane as displayed in Fig. 2.
The actual numbers for the different cut-offs
are given in section 6 and their impact on our
results is discussed there.

\Section{Numerical Results}

We consider the decay function in the form  (see eq. (\ref{3.39}))
\begin{eqnarray}
F(s_a,s_b,s_c)&=&{P}(s_a,s_b,s_c)+{U}(s_a)+{V}(s_b)+{V}(s_c) \nonumber\\
& & +(s_a-s_c){W}(s_b)+(s_a-s_b){W}(s_c) .
\label{res1}
\end{eqnarray}
Here, $s_a,s_b$ and $s_c$ are the kinematical variables defined earlier.
$P$ is a polynomial and  $U,  V$ and $W$ correspond to the $I=0,2$
and $1$  pion rescattering channels. They satisfy dispersion
relations, the generalized Khuri-Treiman equations,
which describe the unitary
corrections to the  $\eta\to 3\pi$ decays (see eqs.
(\ref{3.47},\ref{3.48},\ref{3.49})). We solve the dispersion
relation iteratively, starting from an aproximative solution
obtained by Neveu and Scherk.

Using the existing
one-loop results of Gasser and Leutwyler as basis and
casting their unitary corrections into a dispersion relation,
we are lead to a three times subtracted
dispersion relation for the three amplitudes
$U, V, W$. However, in order to assess more clearly the relevance of our
subtraction procedure, it is useful to consider also the twice subtracted
case where the subtraction polynomial is simply taken as the current algebra
result for the amplitude. We will therefore discuss both forms of the
dispersion relations, starting with the three times subtracted one.

\subsection{Uncertainties}

We can distinguish several sources of uncertainties. The
first is associated with the ambiguities inherent in the
dispersion relations.
These relations do not fix the subtraction points, except of course
for the requirement that they should lie in a region where the
approximate theory makes sense.
In addition, there is the ambiguity to the solution of the Khuri-Treiman
equations related to the asymptotic behaviour of the $\pi\pi$-phase shift
as discussed in sect. 4.
Then, there are the errors in the
input parameters, that is in the values of the subtraction
constants. These uncertainties are of course not completely
disconnected; for instance the one-loop result from which we
determine the subtraction constants may be rather
accurate for some values of the subtraction points but not
for others, and so a wrong choice of the subtraction point
may underestimate the error completely.
And finally, there are technical problems, such as the
convergence of the iteration and the numerical integration in
the complex plane.

\vspace*{1.0cm}
\noindent
1. Subtraction points

We consider first the variation in the subtraction
points.
In section 4, we constructed the second order
subtraction polynomial which was obtained by writing
the one-loop result in the form of a three times subtracted
dispersion relation. This polynomial then serves as the
starting point for the complete dispersion relation. As
discussed in section 4, the choice of the three
subtraction points $s_1, s_2, s_3$ for $U$ and those
for $V$ and $W$, which will be denoted
collectively by $SP$, affects the
subtraction polynomial considerably and consequently the
complete amplitude. Whereas  in chiral perturbation theory and
in view of the treatment in section 4 it seems most natural to subtract
at small $SP$ (at $s_i\approx 0$), varying the points
gives us a feeling for the error.
We have therefore
calculated some important quantities for several $SP$; the
results are shown in Tables 3 and 4. As subtraction points we have taken
the sets in Table 1.

In Table 3 we give the splitting of the one-loop amplitude
(\cite{GL85c}) in the center of the Dalitz plot
into a polynomial and a dispersive part,
as described in eq. (\ref{Pfix}). Furthermore, we list the first
approximation, $\Ph_{10}$, and our final result.
Whereas, by construction
of the subtraction polynomial, the one-loop amplitude at
the center of the Dalitz plot is reproduced for all $SP$,
the relative contributions of the two pieces vary strongly.
However, as long as the subtraction points are not too close to
the two-pion threshold, the total amplitude given in the last column of
Table 3 is rather stable. We therefore consider it most natural to select
a small subtraction point. Excluding $SP=3$, we obtain
$A_{\rm tot}(s_0,s_0,s_0)=1.57\pm 0.12+i(0.41\pm 0.03)$.

Table 4 contains the physical observables for the
various subtraction points.
The rates in the second last two columns are calculated
from eq. (\ref{betamp}), e.g. with the
normalization of the amplitude given by $m^2_1$. We will
discuss the relevance
to the quark masses in the next section.

As may be expected, large subtraction points (near the physical
region) yield smaller rates because there the
full result is nearer to the one-loop result which is
too small. On the other hand, large negative
$SP$ do not  arbitrarily increase the rates; rather, these remain
remarkably stable. Thus it is not possible to (artificially)
enhance the rates by playing with the subtraction points.
The uncertainty of the subtraction point can
be substantially reduced, if the
Dalitz plot slopes $a,b,c$ (to be discussed below)
could be measured with two percent accuracy; This
would allow to narrow down the error bar on $\Gamma$
to 10 \% or 16 eV. At present, the experimental values (which
admittedly have larger errors)  prefer small $SP$.

The values of the physical quantities which correspond
to $SP=0$ will be taken as the central values.

\vspace*{0.5cm}
\noindent
2. Errors in the subtraction constants

Without a next order calculation, the error on the subtraction
constants in the
polynomial, in particular on $\gamma$ and $\delta$ which vanish at tree
level cannot be truly assessed. A reasonable estimate of the errors
can be obtained by the following
observation by Anisovich and Leutwyler
\cite{Heiri}. The current algebra result for
the $\eta\rightarrow 3\pi$ amplitude has an Adler zero
(a value of the kinematic variables where the amplitude
vanish), which for
finite quark masses is shifted to $s\equiv s_A={4\over 3} m_\pi^2$.
This can be easily seen from the tree level result; obviously, the
Adler zeroes lie on a straight line.
The one-loop amplitude shares this feature (with a slightly
shifted $s_A$), if in addition we fix
$s_a=s_c=s_A$ (or $s_a=s_b=s_A$). Moreover, along the
line $s_a=s_c$ the slope
${d A\over ds}(s_a=s_c=s_A)$ is practically unchanged compared to the
leading order expression ${3\over m_\eta^2-m_\pi^2}$. This suggests
that the amplitude along the line $s_a=s_c$ near $s_A$ is very stable
against corrections and that this kinematical point is therefore well
suited as subtraction
point. Note that it is crucial to stay on the line $s_a=s_c$. Consider
for example the subtraction polynomial in the form corresponding
to an expansion around the point $s_a=0$, $s_b=s_c$,
\begin{equation}
P=\alpha+\beta s_a+\gamma s_a^2+\delta (s_b-s_c)^2 .
\label{Psymm}
\end{equation}
The coefficients $\alpha$, $\beta$ in (\ref{Psymm}) are very different
from the current algebra expressions $\alpha^{\rm CA}=
-4 m_\pi^2/(m_\eta^2-m_\pi^2)$,
$\beta^{\rm CA}=3/(m_\eta^2-m_\pi^2)$. The amplitude is not stable
against corrections along the line $s_b=s_c$ and it is difficult to assign
an error bar to $\alpha$, ...$\delta$ in (\ref{Psymm}) as noted before.

We therefore expand around $s_a=s_c=s_A$, i.e.
\begin{equation}
P=\bar b z+\bar c z^2 +\bar d w+\bar e w^2 + \bar f w z ,
\label{PAdler}
\end{equation}
where
\begin{equation}
 z=3 s_0-2 s_A-s_b, \qquad\qquad w=s_a-s_c .
\end{equation}
Due to Bose symmetry only three of the five constants in (\ref{PAdler})
are independent; for instance
\begin{eqnarray}
\bar e&=&{\bar b-\bar d\over 6 (s_0-s_A)}+\bar c \nonumber \\
\bar f&=&{\bar b-\bar d\over 2 (s_0-s_A)}+2 \bar c
\end{eqnarray}
Numerically we find
\begin{eqnarray}
\bar b&=&4.87 {\rm GeV}^{-2} \nonumber \\
\bar c&=&13.65 {\rm GeV}^{-4} \nonumber \\
\bar d&=&11.88 {\rm GeV}^{-2} .
\label{numval}
\end{eqnarray}
Here, only those contributions from the subtraction polynomial
which come from $\bar A^{\rm rem}$ (see section 4.2) have been included.
The remainder $R(s_a,s_b,s_c)$ serves only to bring the one-loop
amplitude into the form of a dispersion relation and is cancelled once
the first iteration step has been performed.
As discussed above, the constants $s_A$ and $\bar b$ are known with very good
accuracy. However, $\bar c$ receives contributions only at one-loop order
and $\bar d$ is very sensitive to corrections.
For the error estimate, we therefore assume that $s_A$ and $\bar b$ are given
exactly
by (\ref{numval}) whereas $\bar c$ and $\bar d$ are assigned a relative error
of 25 \%,
typical of higher loop-corrections:
\begin{eqnarray}
\delta \bar c&=&3.41 {\rm GeV}^{-4} \nonumber \\
\delta \bar d&=&2.97 {\rm GeV}^{-2} .
\label{errorbar}
\end{eqnarray}
We can now translate these errors into uncertainties of the
constants $\alpha$, ..., $\delta$ in the expansion (\ref{Psymm})
by comparing the two forms for $P$ (again, the contributions of the
remainder $R(s_a,s_b,s_c)$ are not written):
\begin{eqnarray}
\alpha&=&-1.02\pm 0.01\mp 0.34 \nonumber \\
\beta&=&(20.35\mp 0.72 \pm 5.69) {\rm GeV}^{-2} \nonumber \\
\gamma&=&(-1.38\pm 13.65 \mp 23.72) {\rm GeV}^{-4} \nonumber\\
\delta&=&(6.22 \pm 0.0 \pm 2.63) {\rm GeV}^{-2} .
\label{errsymm}
\end{eqnarray}
The first and second error is induced by the uncertainty in $\bar c$ and
$\bar d$ respectively. The errors induced by $\bar d$ are particularly large,
however they are correlated and cancel completely for the polynomial
(\ref{Psymm}) evaluated at the center of the Dalitz plot.

The induced error bar on the amplitude at $s_a=s_c=s_0$ can be estimated
by calculating the change in $P(s_0,s_0,s_0)$. Only $\delta c$ matters of
course and we find
\begin{equation}
\delta \bar A(s_0,s_0,s_0)\approx \delta P(s_0,s_0,s_0)=
\delta \bar c (s_a-s_A)^2=0.12 .
\end{equation}
This is 8 \% of the full amplitude corresponding to a 16 \% error on
the rate, i.e. 26 eV if $Q=Q_{\rm Dashen}$ is employed. The actual change
of the full amplitude is somewhat larger and summarized in Table 7.

The rate is thus very sensitive to the value of the constant $\bar c$. If its
higher order contributions
turned out to be large and positive, using
its one-loop value would seriously underestimate the rate. We therefore
attempt a more quantitative estimate of the
higher order corrections to $\bar c$ by considering a twice subtracted
dispersion relation, following again Anisovich and Leutwyler \cite{Heiri}.
We start by decomposing the amplitude into
its isospin components, i.e.
\begin{eqnarray}
F(s_a,s_b,s_c)&=&M_0(s_a)+(s_a-s_c) M_1(s_b)+(s_a-s_b) M_1(s_c) \nonumber \\
&&+M_2(s_b)+M_2(s_c)-{2\over 3} M_2(s_a).
\end{eqnarray}
This decomposition is ambiguous for unitarity determines only the singular part
of the amplitudes $M_0$, $M_1$ and $M_2$. A polynomial in $s_a$ can always be
shifted between the isospin amplitudes, a feature which can be used to make
the $I=1,2$ parts in $F$ small, throughout the physical region.
Then, a dispersive
analysis is performed for the function $M_0$ only, which must satisfy the
twice
subtracted relation
\begin{equation}
M_0(s_a)={1\over D(s_a)} \left\{ \alpha_0+\beta_0(s_a-s_A)+
{(s_a-s_A)^2\over \pi} \int {ds' D(s') f_0^0 (s') {2\over 3} \bar M_0(s') \over
(s'-s_A)^2 (s'-s_a-i\epsilon)} \right\}.
\label{ALdisprel}
\end{equation}
For simplicity, we have neglected the small pieces proportional to
$M_1$, $M_2$ under the integral.
The two subtraction constants $\alpha_0$, $\beta_0$ are fixed
by the requirement
\begin{eqnarray}
M_0(s_A)&=&M_0^{\rm 1-loop} \nonumber \\
{d \over ds_a} M_0(s_A)&=&{d\over ds_a} M_0^{\rm 1-loop}(s_A).
\end{eqnarray}
The 1-loop amplitude enters here only to fix the subtraction constants at a
well suited kinematical point, the Adler point. As mentioned, the
corrections at this point are very small so that this procedure
is reasonable. There is
however a further implicit assumption in the method, namely the
choice of
the $I=1,2$ amplitudes. Although this choice affects the total amplitude, say
at the center
of the Dalitz plot, rather substantially, the resulting uncertainty
on $\bar c$ is
not important. The contribution to $\bar c$ can then be written as a
integral over
the discontinuity in the $s_a$ channel:
\begin{equation}
\bar c^{(M_0)}={1\over 4\pi D(s_A)} \int {ds' D(s') f_0^0(s') [\alpha+\beta
(s'-s_A)+
{2\over 3} \bar M_0(s')] \over (s'-s_A)^3 }.
\end{equation}
Solving eq. (\ref{ALdisprel}) by iteration we obtain
$\bar c^{(M_0)}=(7.9+i 0.5)
{\rm GeV}^{-4}$, which has to be compared to the contribution of
$M_0^{\rm 1-loop}$, $c^{(M_0, {\rm 1-loop})}=5.85 {\rm GeV}^{-4}$. Thus
\begin{equation}
\delta \bar c=2.05 {\rm GeV}^{-4}
\end{equation}
which lies perfectly well within the range established above.
However, the sign of the correction (positive) is now well understood.
The reason is that
current algebra underestimates the $I=0$ s-wave rescattering, hence yielding
too small a discontinuity.

We conclude that the actual value of $\delta \bar c$ is higher
than the one-loop
value; and  we will adopt $\bar c^{(M_0)}=(7.9+i 0.5)
{\rm GeV}^{-4}$ as the most likely number.  It shifts the
amplitude at the
center to 1.71 +$i$0.45 (of which 1.47 is in the subtraction
polynomial) or by about 9 \% \footnote{We have taken the
subtraction point $SP=0$.}.

\vspace*{0.5cm}
\noindent
3. Asymptotic behaviour of the $\pi\pi$-phase shift

In sect. 4 we mentioned an ambiguity to the solution of Khuri-Treiman
equations due to the so far unspecified asymptotic behaviour of the
$\pi\pi$-phase shifts. Here we give a quantitative estimate of the
effect on our results. We neglect the $I=1,2$ phase shifts and consider
the approximate solution $F=P+\Ph_{10}$, where $\Ph_{10}$ is given explicitly
in eq. (\ref{5.28}). This approximation is actually very close to our
final numerical solution, c.f. Table 3. Now we study two cases with
distinct asymptotic behaviour of both, phase shift and amplitude:
i) $\delta_0^0(s)\rightarrow 0$ and $Q=0$, i.e. $F(s) \sim s^2$
and ii) $\delta_0^0(s)\rightarrow \pi$ and $Q={\rm const.} \not= 0$, where
we fine tune $Q$ such that the amplitude has the improved asymptotic behaviour
$F \sim s$. In the first case we use the phase shift as employed before,
given in eq. (\ref{Schenk}) and denoted here
by $\delta_0^{0,\ {\rm Schenk}}(s)$.
In case ii) we need a model which guides
the phase to its asymptotic value $\pi$. We take
\begin{displaymath}
\delta_0^0(s) =
\left\{\begin{array}{ll}
\delta_0^{0,\ {\rm Schenk}}(s), & s\leq s_{00} \\
\delta_0^{0,\ {\rm Schenk}}(s) \left({s\over s_{00}}\right)^h, &
s_{00}\leq s\leq \Lambda_h^2  \\
\pi, &  s > \Lambda_h^2 ,
\end{array} \right.
\end{displaymath}
with $s_{00}=(865 {\rm MeV})^2$ and $\Lambda_h^2$ is the scale where the
phase reaches its asymptotic value. We consider two choices $h=1/4$ and
$1/8$ with corresponding scales $\Lambda_{1/4}^2=(2.16 {\rm GeV})^2$ and
$\Lambda_{1/8}^2=(4.87 {\rm GeV})^2$ respectively.

We then calculate the difference $\delta F=F^{(i)}-F^{(ii)}$ at the center
of the physical region for various subtraction points $SP$. The maximal
difference we obtain is $\delta F=0.015-i*0.020$, or 1 \% and 5 \% of the
full amplitude for real and imaginary part respectively. This in turn implies
an error on the rate $\bar\Gamma$ of less than 3 \% or $\approx 4 {\rm eV}$.
The exercise shows that rather different assumptions on the asymptotic
behaviour imply only modest changes of the amplitude in the low energy
region. Here, taking three subtractions pays visibly. The
same estimate for the twice subtracted case yields an uncertainty in the
real part of the amplitude of already $\approx 5 \%$ or
$\approx 17 {\rm eV}$ on the rate.

\vspace*{0.5cm}
\noindent
4. Technical uncertainties

We continue with some remarks on the structure and
convergence of our iteration.
As described in section 5, we distinguish two iterations,
namely the one in powers of the weaker $I=2,1$
rescatterings ($k$-iteration) and the iteration when solving the
dispersion relation ($m$-iteration). In fact, the latter
is an iteration in the $I=0$
rescatterings.  We will discuss some
exemplary numerical results and comment on the  general
trends. We will always use the subtraction at zero with the
corresponding values of the polynomial.

To illustrate the $m$-iteration, we take the
succesive values of the contribution to
pure $I=0$ problem as considered by Neveu and Scherk. Writing
the amplitude $U$ in the form (see eq. (\ref{ukuk}))
\beq U_k(s)=\sum_{1}^{\infty}\it{\Phi}_{mk}(s) \eeq
we find in the center of the Dalitz plot the following numbers:
\begin{equation}
{\rm Re} \Ph_{10}=0.1828 \quad {\rm Re} \Ph_{20}=-0.0631 \quad
{\rm Re} \Ph_{30}=0.0051 \quad {\rm Re} \Ph_{40}=0.0001 \nonumber
\end{equation}
This suggests that the iteration in the $I=0$ scattering phase converges
nearly like a geometrical series with a coefficent of $0.3$. We have found
this behaviour over the whole Dalitz plot and for
all values of $k$. The iteration is therefore
terminated at $m=5$.

Next we consider the $k$-iteration. As example, we take
again the amplitudes $U$, $V$, $W$ at the center. We obtain for the
corrections to the polynomial:
\medskip

$U_0 = 0.1248 + i0.4786$  no $I=1,2$ corrections

$U_1 = -0.0180 + i0.0638$ one  $I=1,2$ iteration

$U =    0.1145   +i0.5431$ total result

$V_1 = 0.0407 -i0.0617$  one $I=1,2$ corrections

$V = 0.0343-i0.0649$   total

$W_1 = -0.0038 - i0.0003$  one $I=1,2$ corrections

$W = -0.0039 + i0.0002$ total $I=1,2$ iteration

Again we see that the convergence of the results is rather good;
consecutive terms decrease by about $10$ to $15$ percent.

To appreciate the degree to which the final result fulfills
unitarity, we consider again the amplitude  at the center
of the Dalitz plot for various stages of the iteration.
Including only the first iteration, $\Ph_{10}$, the value of the
amplitude is $A_{app} = 1.57+i0.5017$. Unitarity
is checked by inserting
this (and the corresponding values for all $s$) value into
the dispersion relation for the
total amplitude  which yields a new value,
say  $A'_{app}$ on the left hand side. Then the relative difference
$d=(A_{app}-A'_{app})/|A'_{app}|$ is calculated.
At the center of the Dalitz plot, we obtain for this deviation
$0.0441-i0.0301$ which indicates that unitarity is fulfilled
to about six percent. Taking now the full amplitude, we get
at the center again $1.57+i0.4128$. The unitarity check now yields
$d=0.0007+i0.0002$ for the deviation, a rather satisfactory result.
We see that the iteration has improved the unitarity drastically,
although the central value of the amplitude is not much changed.

Another result of interest is the quality of the Neveu-Scherk
$\Ph_{10}$ iteration and the relative sizes of the
various isospin amplitudes
$U, V$ and $W$. As seen from the values given before,
$P+\Ph_{10}$ corresponds rather accurately to the final value. This
is accidental. If we just solve the complete
Neveu-Scherk problem, e.g. only include $I=0$ rescatterings,
the result would be $U=0.1248+i0.4786$ and $\Ph_{10}$
would not be a good approximation. On the other hand, the
complete calculation yields $U=0.1145+i0.5431$, $V=0.0343-i0.0065$,
$W=-0.0039+i0.0002$. This shows that the contribution from $V$
is quite substantial, and that $U$ alone does not saturate the
amplitude at all. Nevertheless we note that $|V/U|$ is roughly
about a third, which justifies our perturbative treatment
of $V$ (and $W$).

Our calculation further depends on input parameters which are not
well known. One is the cutoff
of the dispersion integral. Changing it from
$82m_{\pi}^2$ to $164m_{\pi}^2$, the rates vary at most
by $1$ eV and thus the variation is negligible.

Also, the low energy constant $L_3$ \cite{CH2} is
not very precisely known. Varying $L_3$ within the error bars
we find the following ranges
\medskip

${\it\Gamma}_{\eta\ar\pi^{+}\pi^{-}\pi^{0}}:182.1-181.1eV$

${\it\Gamma}_{\eta\ar\pi^{0}\pi^{0}\pi^{0}}:250.9-253.5eV$

$ a:-1.21- -1.12$

$b:0.25-0.23$

$c:0.10-0.09$

for the one-loop value of the constant $\bar c$.

Apart from the changes in $a$, these variations are small. A
more precise measurement of $a$ could therefore restrict $L_3$.

\subsection{Rates and Dalitz Plot Distribution}

Let us turn now to the values of the physical quantities.
If we take the one-loop value for the constant $\bar c$,
the rate for $\eta\ar\pi^{+}\pi^{-}\pi^{0}$,
normalized by $m^2_1$, is
\beq
\label{valut}
\Gamma = (180 \pm 40)\  {\rm eV}.
\eeq
The error given reflects our estimate of the uncertainty of
25 \% of $\bar c$.
On the other hand,
if the improved value for $\bar c$ is used,
the rate becomes
\beq
\label{ivalut}
\Gamma = 209\ {\rm eV}.
\eeq
We estimate the remaining error on
this result to be about $20$ eV. Since the one-loop
value of $\bar c$ is most certainly too small, the
result in eq. (\ref{ivalut}) should be more
reliable than the value in eq. (\ref{valut})
The corresponding values for the decay into
neutral pions are
$(252 \pm 56) {\rm eV}$ and  $295 {\rm eV}$.
Thus, the rates are not subtstantially larger than the
previous results, if the the one-loop value of
$\bar c$ is taken. However, if its improved value is employed by taking into
account the rescatterings through
the Khuri-Treiman equations, the corrections are substantial,
although the result is still considerably below the
experimental one.
The size of the corrections are qualitatively reasonable: since the
one-loop corrections amount to about 50 \% in amplitude, next
order corrections
might be expected at about 40 eV, roughly the value we found.

The ratio
\beq
r=\frac{{\it\Gamma}_{\eta\ar\pi^{0}\pi^{0}\pi^{0}}}
{{\it\Gamma}_{\eta\ar\pi^{+}\pi^{-}\pi^{0}}}
\eeq
between the rates of the neutral and charged pion channel
takes the values
\beq
r=1.40 \pm0.03
\eeq
and
\beq
r=1.41 \pm0.03
\eeq
for the two values of $\bar c$, respectively.
The values correspond to the
the small subtraction points $SP$ considered; the error is obtained
by varying the $SP$ between $-10$ and $2$ (see Table 4).
In comparison, the value
$r=1.43$ was obtained in Ref. \cite{GL85c}.
The particle data group quotes
\begin{eqnarray}
r&=&1.35 \pm 0.05 \qquad {\rm "our\  fit"}  \nonumber\\
r&=&1.27 \pm 0.14 \qquad {\rm "our\  average"},
\end{eqnarray}
favoring smaller values for $r$, in particular almost excluding the current
algebra result.
Recently, $r$ has been remeasured in a
direct measurement \cite{Ams95}
\beq
r=1.44 \pm0.09 \pm0.01
\eeq
with smaller errors than previously and
which is in very good agreement with our result and the one-loop value
but somewhat below the current algebra prediction of $r=1.51$.

The distribution over the Dalitz plot is conventionally described
by the two variables $x$ and $y$
\begin{eqnarray}
 x&=&\frac{\sqrt3}{2m_{\eta} Q_\eta}(s_c-s_b) \\
 y&=&\frac{3}{2m_{\eta}Q_\eta}((m_{\eta}-m_{\pi})^2-s_a) - 1 \\
 Q_\eta&=&m_{\eta}-2 m_{\pi^+}-m_{\pi^0}.
\end{eqnarray}
Since $Q_\eta$ is small, it is important to keep the masses of the charged
and neutral pions different. The mass difference is generated by
electromagnetic corrections. However, as discussed in the first section,
the other electromagnetic contributions to the decay amplitude
are small and can be neglected.

The Dalitz plot distribution $|A(x,y)|^2$
can be parametrized by a second order polynomial.
For the decay into charged and neutral pions we may write, respectively,
\begin{eqnarray}
|A^{+-0}(x,y)|^2&=&N^{+-0}\left(1+ay+by^2+cx^2\right), \nonumber\\
|A^{000}(x,y)|^2&=&N^{000}\left(1+g(x^2+y^2)\right).
\end{eqnarray}
Of course, the neutral pion mass must be used for $x$ and $y$ in the
second expression.

The various determinations of the slopes $a,b,c$ and $g$ are as follows:

Gasser and Leutwyler \cite{GL85c}:  $a=-1.33, b=0.42, c=0.08$

Layter et al. (exp)\cite{Lay74}:
$a=-1.08\pm0.014, b=0.034\pm0.027, c=0.046\pm0.031$

Gormley et al. (exp)\cite{Gor74}:
$a=-1.17\pm0.02, b=0.21\pm0.03, c=0.06\pm0.04$

Amsler et al. (exp)\cite{Ams95}:
$a=-0.94\pm 0.15, b=0.11\pm 0.27$

Alde et al. (exp)\cite{Alde}:
$g=-0.044\pm 0.046$

This work:    $a=-1.16, b=0.24, c=0.09, g=-0.028$ (one-loop value
for $\bar c$)

\hspace{1.9 cm} $a=-1.16, b=0.26, c=0.10, g=-0.014$ (improved value
for $\bar c$).

We see that our numbers are very near to the experimental values, in
fact closer than the one-loop results. The
values of $a$ for negative $SP$ are somewhat too large, while
the result for $SP=0$ correspond nicely to the result
of Gormley et.al. Unfortunately, the two most accurate experiments are mutually
inconsistent and thus preclude a definite statement;
measuring the slopes with larger accuracy would remove the
uncertainties inherent in the choice of the subtraction points.
The quadratic slope
parameter of the neutral decay mode, $g$, is experimentally
compatible with zero. We observe that the slopes (with the
exception of $g$) do not depend
strongly on the value of $\bar c$.

The shape of the amplitude, however, changes compared
to the one-loop amplitude. In Fig. 3 we plot the real part of our
numerical solution (with the onel-loop value of $\bar c$) along the
line $s_a=s_c$, together with the current algebra
prediction as well as the chiral perturbation theory one-loop amplitude.
At small $s$, all amplitudes are close together which reflects the fact
that we have subtracted at $s_a=0$.
The cusp generated by the two-pion threshold is more pronounced in the
solution taking into account the corresponding final state interactions
to all orders. However, the requirement of unitarity bends the amplitude
down more
strongly in the physical region, leading to a value of the amplitude at
the center of the Dalitz plot close to the value obtained to one-loop
ChPT. At the same time, the slope of the amplitude is reduced substantially
which yields a smaller values of the linear slope parameter of the Dalitz plot
distribution. The imaginary part is seen to be enhanced over the whole
physical region, c.f. Fig. 4. This was expected, as the one-loop amplitude
gives only the leading term to the imaginary part. The figures
for the improved $\bar c$ have a similar shape, but with different
normalization corresponding to the larger rate discussed above.

\subsection{Two Subtractions}
We turn briefly to the twice subtracted dispersion relation. Starting
from the current algebra polynomial, we  obtain the results
in Table 6, taking the subtraction points in Table 5. We have
used the same scattering phase as in the three times subtracted
case, and set the polynomial $Q$ equal to zero. This time,
the rates grow fast
if we change the subtraction points to large
negative values. On the other hand, this increase is accompagnied
by an unreasonable change of the slopes $a$ and $b$ as well as
unacceptable values of $r$. We expect similar changes, if
we vary the asymptotic behaviour of the amplitude and the phase.
We conclude
that artificially enhancing the rates with large negative
subtraction points does not yield correct results and that
the higher order terms in the subtraction polynomial are
essential.
An alternative approach would be to use the one-loop subtraction
data,
but maintain a twice subtracted relation. We expect in this
case larger uncertainties from the assumption of elastic unitarity
as well as from the asymptotic ambiguities.

Roiesnel and Truong \cite{RT81} were the first to
invoke unitary corrections to enhance the decay rate. These
authors
obtained satisfactory values for the rates, using a once subtracted
dispersion relation;
furthermore they only considered $I=0$ rescatterings, but only for one part of
the amplitude. As a result, their discontinuity is not
given by the usual form, but larger by the factor $9/5$. Since
the sign of the $I=2$ contribution is negative, this
factor in fact gets enhanced because the $I=0,2$
pieces suffer an artificial cancellation. If this factor is omitted, the
result is not as dramatic and close to our results for
$SP=0$. As noted, the  dispersion relation with one subtraction
is subject to large ambiguities and thus the result will be beset
with high uncertainties.

\Section{Conclusions}

The decay amplitude of $\eta\to 3\pi$ is proportional
to the quark mass
difference  $(m_{d}-m_{u})$ and thus
provides one of the ingredients to determine the important
ratio $\frac{m_{u}}{m_{d}}$. Including also higher orders in chiral
perturbation theory, the decay rate can indeed be written
as
\beq \label{relrate}
\Gamma = ( Q_{DT}/Q)^4 \bar\Gamma
\eeq
where
\beq \label{ququ}Q^{-2}\equiv\frac{m_{d}-m_{u}}
{m_{s}-{\hat m}}
\cdot\frac{m_{d}+m_{u}}{m_{s}+{\hat m}} \nonumber \eeq
with ${\hat m}\equiv\frac{1}{2}(m_{d}+m_{u})$. $Q_{DT}$ is
the value of $Q$ if Dashen's theorem for the electromagnetic
kaon mass difference is used (see eq. (\ref{betamp})) and
$\bar\Gamma$ the corresponding width as given in the previous
chapter.

Using the value $Q_{DT}$ (or the corresponding and
long established values for the quark mass ratios),
the one-loop ChPT prediction for the rate
given by Gasser and Leutwyler \cite{GL85c} has been
considerably below the experimental number.
Theoretically, the rate can be
increased by lowering the ratio
$m_{u}/m_{d}$ \cite{DHWL} or by including further corrections
in $\bar\Gamma$.
In particular,
it has long been suspected that the unitary corrections may
be sufficient to enhance the rate sufficiently.

The existing framework embodied in the so called Khuri-Treiman
equations \cite{KT60} allows to take into account rescatterings of two
pions which are thought to dominate. Some time ago, Roisnel
and Truong \cite{RT81} claimed that in fact these corrections
solve the problem, but as
explained in section 6 (see also \cite{GL85c}),
we believe that their results are an overestimate.

In this paper we have calculated the unitary corrections using
the Khuri-Treiman equations (which include the $I=0,2$ pion pion
rescatterings), complemented with the  $I=1$ interactions.
To specify these dispersion
relations completely, that is to define the subtraction
method, we have used an iterative procedure for solving it
which accounts in lowest order for the
one-loop ChPT results of
Gasser and Leutwyler. This forced us to a three times
subtracted dispersion relation with rather fast convergence
and little dependence on the unknown high energy scattering phase.
In this approach, the subtraction polynomial is quadratic in the
invariant momenta, rather than the linear form of current algebra.
The major source of errors lies in the choice
of the subtraction points and, particularly, in the
uncertainties in the subtraction
constants.  The constants connected to the
quadratic terms vanish at tree level and only
start at the one-loop level; thus their one-loop value is rather uncertain.
In section 6 we have
described how this problem can be overcome (following recent
work of Anisovich and Leutwyler \cite{Heiri}), and
have given a reasonable estimate of the constant $\bar c$. We have chosen
to give the results corresponding to both their one-loop value and to
the improved one. Using everywhere the new value of $f_\pi$ \cite{fpi},
the rate corresponding to the the one-loop value is
\beq
\label{rresult}
\bar\Gamma = (180\pm 40) {\rm eV},
\eeq
while the improved number of  $\bar c$ yields
\beq
\label{newresult}
\bar\Gamma =209 {\rm eV}
\eeq
and where we estimate the remaining errors to be about $20 eV$.
We view the second result as the more reliable one.
Although the corrections are quite
large, they do not suffice to explain the experimental value of $(281\pm 28)
{\rm eV}$.

 From eq. (\ref{newresult}) we obtain
\beq
Q = 22.4 \pm 0.9.
\eeq
while the rate of $180$ eV in (\ref{rresult}) would imply $Q = 21.6 \pm 1.3$.
In contrast,  $Q_{DT}=24.1$.

These lower values of $Q$ can obviously  be accounted for if the
ratio
$\frac{m_{u}}{m_{d}}$ is lowered, and another mass ratio, say
$\frac{m_{d}+m_{u}}{m_{s}}$
is kept fixed.  In this case, $\frac{m_{u}}{m_{d}}$ is
reduced from $0.57$ to $0.49\pm 0.04$ or to
$0.52\pm 0.03$ depending on whether eq. (\ref{rresult}) or
eq. (\ref{newresult}) is used.
Thus, while it may be somewhat smaller than previously
thought, the up quark mass does not vanish \footnote{
our analysis is based on the assumption of the
validity of standard chiral perturbation theory. In contrast, a
treatment along the lines of Ref. \cite{GCHPT} might lead to
different results; however the larger quark masses in that scheme
exclude a zero mass automatically}. We note that the somewhat lower value
of $\frac{m_{u}}{m_{d}}$ in \cite{DHWL}
was obtained with a changed value of $\frac{m_{d}+m_{u}}{m_{s}}$.
The constant $Q$ can also be determined from electromagnetic
corrections to the meson masses.
As already noted in \cite{DHWL}, these lower values of $Q$ correspond
roughly
to the one obtained from the kaon mass difference, if electromagnetic
corrections are positive and large as found in \cite{DHW}.

We have also determined the slopes in the
Dalitz plot distributions; the results are given in
the last section. We find that the ratio $r$ between the rates
of the neutral and charged pion mode remains roughly at
1.4 which is also favored experimentally \cite{Ams95}. On the
other hand, the
slopes change from their one-loop ChPT values; our
number for $a$ is higher than
the previous result and nearer to the experimental value.

The experimental situation is rather unsatisfactory.
The decay rates are normalized with respect to the decay with
$\Gamma(\eta\rightarrow \gamma\gamma)$, where the results from
two-photon production disagree with those from Primakoff production.
The two-photon measurements seem more reliable; however, in order
to resolve the issue completely a reanalysis of the Primakoff data
would be necessary. The Dalitz plot distribution of $\eta\rightarrow
\pi^+\pi^-\pi^0$ has been measured with rather high accuracy
\cite{Gor74,Lay74}. However, the assumptions made by these authors are
not compatible and do not allow comparison ot the numerical values. A
recent experiment \cite{Ams95} has still too large error bars in order
to be conclusive.
As our results show, there is a rather strong
correlation between rates and slope parameters and a more
acurate measurement of the latter would fix the rates, and
thereby the up quark mass better.

\vspace{1.0cm}
\noindent
{\bf Acknowledgments:}
We benefited much from discussions with J. Bronzan, B. Holstein, J. Gasser,
A. Salathe,
J. Stern, J. Donoghue and, in particular, H. Leutwyler who shared with us
much of his recent results prior to publication.
Large part of the work by J.K. was done while at
the Department of Physics and Astronomy, University of Massachusetts, Amherst;
J.K. and D.W. also thank the Institute for Nuclear Theory in Seattle
where some of the work was done for hospitality.

\appendix\def\theequation{\Alph{section}.\arabic{equation}}
\def\Section*#1{\setcounter{equation}{0}\section{#1}}
\renewcommand{\thesection}{}
\Section*{Appendix A}
In this Appendix we give the three projections of the
subtraction polynomial $P$. Since $P$ has no cuts, the
integrals as defined in section 3 are trivial
and may be performed explicitly.

The second order subtraction polynomial is of the form
\beq P(s_{a},s_{b},s_{c})=
\al+\be s_{a}+\ga s_{a}^{2}+\de(s_{b}-s_{c})^{2}
\eeq
and we need the projections
${\overline{s_{b}^{n}}}$ for $n=0,..,3$. The
corresponding results are
\beqq {\overline 1}&=&1 \nonumber \\
{\overline{s_{b}}}&=&\frac{3s_{0}-s_{a}}{2} \\
{\overline{s_{b}^{2}}}&=&
\left(\frac{3s_{0}-s_{a}}{2}\right)^{2}
+\frac{K^{2}(s_{a})}{3} \nonumber \\
{\overline{s_{b}^{3}}}&=&\frac{3s_{0}-s_{a}}{2}
\cdot\left[\left(\frac{3s_{0}-s_{a}}{2}\right)^{2}
+K^{2}(s_{a})\right]. \nonumber \eeqq
We now insert $P$ into eq. (\ref{3.17}),
reexpress $s_{c}=3s_{0}-s_{a}-s_{b}$ and obtain
\beqq
{\overline P}(s_{a})&=&\al+\be s_{a}+\ga s_{a}^{2} \nonumber \\
&+&\de(4{\overline{s_{b}^{2}}}-4{\overline{s_{b}}}(3s_{0}-s_{a})
+(3s_{0}-s_{a})^{2}) \\ &=&\al+\be s_{a}+\ga s_{a}^{2}
+\de\frac{4K^{2}(s_{a})}{3}. \nonumber \eeqq
For ${\tilde P}(s_{a})$ we get from eq. (\ref{3.18})
\beqq
{\tilde P}(s_{a})&=&\al+\be {\overline{s_{b}}}
+\ga {\overline{s_{b}^{2}}} \nonumber \\
&+&\de({\overline{s_{b}^{2}}}
-2{\overline{s_{b}}}(3s_{0}-2s_{a})+(3s_{0}-2s_{a})^{2}) \\
&=&\al+\be\frac{3s_{0}-s_{a}}{2}-\de s_{a}(3s_{0}-2s_{a})
\nonumber \\
&+&(\ga+\de)\left[\left(\frac{3s_{0}-s_{a}}{2}\right)^{2}
+\frac{K^{2}(s_{a})}{3}\right].
\nonumber \eeqq
Turning finally to ${\hat P}$ its
insertion into eq. (\ref{3.29}) yields
\beqq
{\hat P}(s_{a})&=&-\frac{3}{4K^{2}(s_{a})}
[\,2\{\al{\overline{s_{b}}}
+\be {\overline{s_{b}^{2}}}+\ga {\overline{s_{b}^{3}}} \nonumber \\
&+&\de({\overline{s_{b}^{3}}}-2{\overline{s_{b}^{2}}}(3s_{0}-2s_{a})
+{\overline{s_{b}}}(3s_{0}-2s_{a})^{2}\}
\nonumber \\
& &-(3s_{0}-s_{a})\{\al+\be {\overline{s_{b}}}
+\ga {\overline{s_{b}^{2}}}  \\ &+&
\de({\overline{s_{b}^{2}}}
-2{\overline{s_{b}}}(3s_{0}-2s_{a})
+(3s_{0}-2s_{a})^{2})\}] \nonumber \\
&=&-\frac{1}{2}\left[\be
+(\ga-\de)(3s_{0}-s_{a})\right].
\nonumber \eeqq

\appendix\def\theequation{\Alph{section}.\arabic{equation}}
\def\Section#1{\setcounter{equation}{0}\section{#1}}
\renewcommand{\thesection}{}

\Section{Appendix B}
\setcounter{section}{2}

In this Appendix we discuss the parametrization of the
different pion-pion phase shifts used in our numerical
computations and disply the resulting Omn$\grave{\mbox{e}}$s function.

A careful analysis of the different restrictions on the
phase shifts is carried out in Ref. \cite{AS}. There are two
 main points to be respected, namely first the
threshold behaviour of Re$f_{l}^{I}$
\beq {\mbox{Re}}f_{l}^{I}(s)=\roo^{l}\left(a_{l}^{I}
+b_{l}^{I}\roo+...\right) \eeq
where we used the notations of section 3 and where
$a_{l}^{I}$ denote the different scattering lengths,
$b_{l}^{I}$ the corresponding slope parameters.
Second, one has to implement that the phase shifts
pass through $\frac{\pi}{2}$ at some experimentally
known values $s=s_{l}^{I}$ of the energy. A simple
parametrization respecting those conditions is given by \cite{AS}
\beqq
\label{Schenk}
& &\tan\de_{l}^{I}(s)=\sqrt{\frac{s-4}{s}}\cdot\roo^{l}\cdot
\frac{s_{l}^{I}-4}{s_{l}^{I}-s} \\ & &\quad\cdot
\left(a_{l}^{I}+{\tilde b}_{l}^{I}
\roo+c_{l}^{I}\roo^{2}\right) \nonumber \eeqq
where the threshold expansion is reproduced with
\beq {\tilde b}_{l}^{I}=b_{l}^{I}-a_{l}^{I}\frac{4}{s_{l}^{I}-4}
+(a_{l}^{I})^{3}\de_{l0}. \eeq
As the experimental data on pion-pion scattering near
threshold are rather poor one uses the results of
ChPT for the scattering length \cite{GL85a,GL83}
\beqq & & a_{0}^{0}=0.20,\quad a_{0}^{2}
=-0.042,\quad a_{1}^{1}=0.037, \\
& & b_{0}^{0}=0.24,\quad b_{0}^{2}=-0.075.
\nonumber \eeqq
The remaining data are extracted from experiment
\beqq & & b_{1}^{1}=0.005, \nonumber \\
& & c_{0}^{0}=0,\quad c_{0}^{2}=0,\quad c_{1}^{1}=0,
\\ & & s_{0}^{0}=38.45,\quad s_{0}^{2}=-24.11,
\quad s_{1}^{1}=30.39 \nonumber \eeqq
where $s_{l}^{I}$ is given in units of $m_{\pi}^{2}$ as usual.
These values correpond to the 'best fits' in \cite{AS}.
We remark that the constraints of the Roy equation
around $s=4$ are taken into account in this parametrization.

We already noted that all the numerical
integrals will be cut off. In order to avoid
discontinuities in the numerical integrations,
caused by the step-function induced from the cut-off,
we bring the phases smoothly down to zero multiplying
them with the exponential suppression factor
\beq e^{-k\cdot(s-s_{0}^{0})^{4}} \eeq
ensuring differentiability at $s_{0}^{0}$.
For $k$ we chose the value $k=10^{-5}$.

We do not give explicitely the numerical result for the
Omn$\grave{\mbox{e}}$s function. We just note that instead of
working directly with the definition in eq. (\ref{5.10})
we prefer to go over to the once subtracted form
\beq
D(z)= e^{-\frac{z}{\pi}\intgr ds'
\frac{\de^{0}_{0}(s')}{s'(s'-z)}}.
\eeq
This is possible as we are dealing only with fractions
of such functions in which the extra contribution from the
subtraction point at $z=0$ cancels. The advantage of subtracting
here is the better numerical convergence of the integrals.

\appendix\def\theequation{\Alph{section}.\arabic{equation}}
\def\Section*#1{\setcounter{equation}{0}\section{#1}}
\renewcommand{\thesection}{}

\Section*{Appendix C}
\setcounter{section}{3}
In this Appendix we analyse possible consequences of the
singularities
occuring in the projected amplitudes as displayed in
eqs. (\ref{3.37}) and (\ref{3.38}). We show in particular
that for physical values of the energy, $s+i\ep$, the
amplitude resulting from the dispersion integrals is
finite.

Within the iterative procedure we have to perform dispersion
integrals
over projected functions having poles at $s=(m-1)^{2}$
to obtain certain contributions to the amplitudes in
question. As long as the value of the external variable
$s$ is not around $s=(m-1)^{2}$ there are no problems
with the integrations. We have  thus to analyse only
the case
$s\sim (m-1)^{2}$ where the two types of integrals
are of the form
\beqq \label{a3.1} & &\mbox{const.}
\cdot\int_{(m-1)^{2}-\de-g}^{(m-1)^{2}-\de+g} ds'
\frac{1}{s'-(m-1)^{2}+\de-i\ep}\cdot\frac{1}
{((m-1)^{2}-s')^{\frac{1}{2}+n}}
\nonumber \\ & &\quad\quad\quad\quad\quad\quad+
\mbox{finite}.
\eeqq
Above we introduced the two positive constants
$g>\de>0$
dealing thus with the case $s=(m-1)^{2}-\de\ar
(m-1)^{2}$
from below. The other case is then treated similarly.
$g$
should be small in the sense that we may replace
the full
integrand by its above reduction to the two
rational functions.
'finite' denotes the obviously finite rest and
$n$ the two cases
coming from eq. (\ref{3.37}) with $n=0$ and from
eq. (\ref{3.38}
with $n=1$. After a change of variable $s'\ar
x\equiv s'-(m-1)^{2}+\de$
we obtain
\beq I_{n}\equiv\int_{-g}^{+g} dx
\frac{1}{x-i\ep}\cdot\frac{1}{(-x+\de)^
{\frac{1}{2}+n}}\eeq
discarding all unimportant contributions to
eq. (\ref{a3.1}).
Using the Sokhotsky-Plemelj formula we obtain
the principal value
integral
\beq I_{n}= i\pi\frac{1}{\de^{\frac{1}{2}+n}}+
P\int_{-g}^{+g} dx \frac{1}{x}\cdot\frac{1}
{(-x+\de)^{\frac{1}{2}+n}}.\eeq
As the square-root has a cut we have to distinguish the
cases $-x+\de>0$ and $-x+\de<0$.

For $-x+\de>0$ we obtain
\beq \int dx \frac{1}{x}\cdot\frac{1}
{(-x+\de)^{\frac{1}{2}+n}}
=\frac{2}{\de^{\frac{1}{2}+n}}\int dz\frac{1}{z^{2}-1}
\cdot\frac{1}{z^{2n}} \eeq
making the substitution $z\equiv
\sqrt{\frac{-x+\de}{\de}}$.
For $n=0$ this leads to the contribution
\beq -\frac{1}{\de^{\frac{1}{2}}}\log
\frac{\sqrt{\frac{g+\de}{\de}}-1}
{\sqrt{\frac{g+\de}{\de}}+1}, \eeq
for $n=1$ to
\beq -\frac{1}{\de^{\frac{3}{2}}}\left\{2\cdot
\sqrt{\frac{\de}{g+\de}}
+\log\frac{\sqrt{\frac{g+\de}{\de}}-1}
{\sqrt{\frac{g+\de}{\de}}+1}
-2\cdot\lim_{x\ar\de}\sqrt{\frac{\de}{-x+\de}}\right\}
\eeq
where we properly distinguished the two
cases $|z|<1$ and
$|z|>1$ in the range of integration and where
the principal
value prescription leads to the cancelation
of the infinities at $x=0$.

For $-x+\de<0$ we have to continue analytically.
With $\sqrt{-x+\de}\equiv i\sqrt{x-\de}$ we find
\beq \int dx \frac{1}{x}\cdot\frac{1}{(-x+\de)^
{\frac{1}{2}+n}}
=(-)^{n}\frac{-2i}{\de^{\frac{1}{2}+n}}\int dz
\frac{1}{z^{2}+1}\cdot\frac{1}{z^{2n}} \eeq
making the substitution $z\equiv\sqrt{\frac{x-\de}{\de}}$.
For $n=0$ this leads to the contribution
\beq -\frac{2i}{\de^{\frac{1}{2}}}
\arctan\sqrt{\frac{g-\de}{\de}}, \eeq
for $n=1$ to
\beq -\frac{2i}{\de^{\frac{3}{2}}}\left
\{\sqrt{\frac{\de}{g-\de}}
+\arctan\sqrt{\frac{g-\de}{\de}}-\lim_{x\ar\de}
\sqrt{\frac{\de}{x-\de}}\right\} .\eeq
Here no further distinctions have to be made.

We collect all the contributions for $n=0$
\beq I_{0}=\frac{1}{\de^{\frac{1}{2}}}
\left\{i\pi-\log\frac{\sqrt{\frac{g+\de}
{\de}}-1}{\sqrt{\frac{g+\de}{\de}}+1}-2i
\arctan\sqrt{\frac{g-\de}{\de}}\right\} \eeq
and for $n=1$
\beqq I_{1}&=&\frac{1}{\de^{\frac{3}{2}}}
\{i\pi-2\cdot\sqrt{\frac{\de}{g+\de}}
-\log\frac{\sqrt{\frac{g+\de}{\de}}-1}
{\sqrt{\frac{g+\de}{\de}}+1}
-2i\sqrt{\frac{\de}{g-\de}} \\ & &-2i
\arctan\sqrt{\frac{g-\de}{\de}}
+2\cdot\lim_{x\ar\de}\sqrt{\frac{\de}{-x+\de}}
+2i\lim_{x\ar\de}\sqrt{\frac{\de}{x-\de}}\}.
\nonumber \eeqq
In the limit $\de\ar 0$ a Taylor
expansion shows that the log-terms
vanish whereas the arctan-terms
cancel the $i\pi$-contributions such
that $I$ remains finite.

In the case $n=1$ there remains the
divergent contribution
$\sim\,\sqrt{\frac{\de}{x-\de}}$.
Resubstitution of $x=s-(m-1)^{2}+\de$ yields
a square-root singularity at $s=(m-1)^{2}$.
As this type of behaviour occurs in the
function $W(s_{a})$ only which is always
multiplied with $s_{b}-s_{c}=2\cos
\th_{ba}\cdot K(s_{a})$ we see that the
square-root singularity is lifted by the
Kacser function $K$. The case
$s=(m-1)^{2}+\de$ leads to the same results in the
limit $\de\ar 0$. The resulting amplitudes are
thus finite at $s=(m+1)^{2}+i\ep$ although
discontinuities may occur at this point.

At the unphysical boundary $s=(m+1)^{2}-i\ep$
the amplitudes badly diverge as the
arctan contribution does
not cancel anymore the $i\pi$-terms as above.

\appendix\def\theequation{\Alph{section}.\arabic{equation}}
\def\Section*#1{\setcounter{equation}{0}\section{#1}}
\renewcommand{\thesection}{}
\Section*{Appendix D}
\setcounter{section}{4}
Here we give the explicit form of the functions $R^{(I)}$ occuring in
the determination of the subtraction polynomial for finite subtraction
points, c.f. section 4, eq. (\ref{Rfinite}). The derivation employs the
identity
\begin{eqnarray}
{\rm Re}\, \tilde\Delta_I (s)&=& (s-s_i)(s-s_j) y(s,s_i,s_j) \nonumber\\
&&+ {s\over 2} \left[ s_i \left( y(0,s_i,s_j)+y(0,0,s_i) \right)
                        + (i \leftrightarrow j) \right] \nonumber \\
&& +s_i s_j y(0,s_i,s_j) ,
\end{eqnarray}
where
\begin{equation}
\tilde \Delta_I(s)={s^2\over \pi} \int_{4 m_\pi^2}^\infty ds'
{\delta^{I,{\rm ChPT}}(s')\over s'^2 (s'-s-i\epsilon)}
\end{equation}
and
\begin{equation}
y^I(s,s_i,s_j)={1\over \pi} P \int_{4 m_\pi^2}^\infty ds'
{\delta^{I,{\rm ChPT}}(s')\over (s'-s)(s'-s_i)(s'-s_j)} .
\end{equation}
$P$ denotes the principal value and $\delta^{I,{\rm ChPT}}$ is the
$\pi\pi$-phase with isospin $I$ to leading order in ChPT. For $I=0,2$ we
thus find
\begin{eqnarray}
R^{(I)}(s;s_1,s_2,s_3;s_A)&=& {(s_A-s)\over 3}\sum_{i<j}^{1,2,3}
\left\{ {s\over 2} \left[ s_i \left( y^I(0,s_i,s_j)+y^I(0,0,s_i) \right)
                        + (i \leftrightarrow j) \right] \right. \nonumber \\
&& \qquad\qquad\qquad \left. -s_i s_j y^I(0,s_i,s_j) \right\} \nonumber\\
&&+y^I(s_1,s_2,s_3){1\over 3} \sum_{i<k, k\not=i,j}^{1,2,3}
(s-s_i)(s-s_j)(s_A-s_k).
\end{eqnarray}
The corresponding expression for $I=1$ is
\begin{eqnarray}
R^{(1)}(s_a,s_b,s_c;w_1,w_2)&=& \left({9\over 8}s_0^2-{3\over 2}s_a s_0
+{3\over 8}s_a^2-{1\over 8}(s_b-s_c)^2\right) \nonumber\\
&& \times \left[ w_i \left( y^1(0,w_1,w_2)+y(0,0,w_1) \right)
                        + (1 \leftrightarrow 2) \right] \nonumber \\
&&+{3\over 2} (s_a-s_0) w_1 w_2 y^1(0,w_1,w_2) .
\end{eqnarray}

\mbox{}
\newpage

 {\LARGE\bf Tables}

\mbox{}

\mbox{}

\mbox{}


\begin{center}
\begin{tabular}{|l|l|l|l|}
\hline
SP    &   $s_1$    &     $s_2$     &    $s_3$ \\
\hline
-10   &   $-9.6$   &    $-9.8$     &  $-10.0$ \\
\hline
-6  &     $-5.6$   &    $-5.8$     &  $-6.0$ \\
\hline
-4   &   $-3.5$   &    $-3.7$     &  $-3.9$ \\
\hline
-2  &     $-1.7$   &    $-1.9$     &  $-2.1$ \\
\hline
0   &   $5*10^{-5}$   &    $1.2*10^{-3}$     &  $5*10^{-3}$ \\
\hline
1   &   $1.2$   &    $1.33$     &  $1.4$ \\
\hline
2   &   $1.7$   &    $1.9$     &  $2.1$ \\
\hline
3  &     $2.6$   &    $2.7$     &  $2.8$ \\
\hline
\end{tabular}

\begin{quote}
{Table 1: Values of the subtraction points $SP$
used in the three times subtracted dispersion
relation in units of
 $m_{\pi}^2$. The subtraction points are the
same for $U$ and for $V$, for $W$ we take the first two
values on each row (see also text).}
\end{quote}
\end{center}

\mbox{}

\begin{center}
\begin{tabular}{|l|l|l|l|l|}
\hline
$SP$   & $\alpha$  & $\beta$  & $\gamma$ & $\delta$ \\
\hline
$-10$  & $-1.38$   & $19.73$  & $-7.87$  &  $3.74$ \\
\hline
$-6$   & $-1.28$   & $20.52$  & $-5.45$  &  $4.00$ \\
\hline
$-4$   & $-1.25$   & $20.97$  & $-3.45$  &  $4.13$ \\
\hline
$-2$   & $-1.27$   & $21.39$  & $-1.90$  &  $4.22$ \\
\hline
$0$    & $-1.28$   & $21.81$  & $4.09$   &  $4.19$ \\
\hline
$1$    & $-1.34$   & $21.99$  & $11.24$   &  $3.88$ \\
\hline
$2$    & $-1.38$   & $21.90$  & $17.46$   &  $3.44$ \\
\hline
$3$    & $-1.47$   & $21.08$  & $36.67$   &  $1.81$ \\
\hline
error bar & $\pm 0.14$ & $\mp 1.52$ & $\pm 3.18$ & $\mp 1.08$\\
\hline
\end{tabular}

\mbox{}

\begin{quote}
{Table 2: Coefficients of the subtraction polynomial
for the sets of subtraction points specified in Table 1. The last line
gives the error bar due to the uncertainty in the determination of $L_3$.
All values given here and in the subsequent tables
refer to the one-loop value of the constant
$\bar c$ discussed in section 6.}
\end{quote}
\end{center}

\mbox{}


\begin{center}
\begin{tabular}{|l|l|l|l|l|}
\hline
SP    &   $P$&$ReGL$ &  $A_{app}$& $A_{tot}$    \\
\hline
-10   &   $0.86$   &    $0.64$  &$1.62+i0.44$   &$1.63+i0.41$ \\
\hline
-6  &     $1.10$   &    $0.40$  &$1.70+i0.49$   &$1.69+i0.44$ \\
\hline
-4  &     $1.21$   &    $0.29$  &$1.69+i0.50$   &$1.67+i0.44$  \\
\hline
-2   &    $1.26$   &    $0.24$  &$1.60+i0.49$   &$1.59+i0.42$ \\
\hline
0  &      $1.40$   &    $0.10$  &$1.57+i0.50$   &$1.57+i0.41$ \\
\hline
1 &       $1.46$   &    $0.11$  &$1.51+i0.40$   &$1.50+i0.39$ \\
\hline
2 &       $1.49$   &    $0.01$  &$1.42+i0.49$   &$1.46+i0.38$ \\
\hline
3 &       $1.58$   &    $-0.08$ &$1.32+i0.48$   &$1.41+i0.36$ \\
\hline
\end{tabular}

\begin{quote}
{Table 3: The polynomial approximation P and the real
part of the dispersive part of the one-loop
result, ReGL, in the center of the Dalitz
plot. The sum is fixed to give 1.5, the value
of the one-loop amplitude. Also the result of the
first approximation (including only
$\Ph_{10}$)
and the total amplitude
are given}
\end{quote}
\end{center}

\mbox{}

\begin{center}
\begin{tabular}{|l|l|l|l|l|l|l|l|}
\hline
SP    &$a$&$b$&$c$&$g$ & ${\it\Gamma}_{+-0}$ &${\it\Gamma}_{000}$ & $r$ \\
\hline
-10   &$-1.27$&$0.32$ &$0.07$&-0.023 &$201eV$&$278eV$&$1.38$\\
\hline
-6    &$-1.22$&$0.29$ &$0.08$&-0.020 &$213eV$&$297eV$&$1.40$\\
 \hline
-4    &$-1.21$&$0.30$ &$0.08$&-0.017 &$211eV$&$295eV$&$1.40$\\
\hline
-2    &$-1.21$&$0.30$ &$0.08$&-0.020 &$189eV$&$263eV$&$1.40$\\
\hline
0     &$-1.16$&$0.24$ &$0.09$&-0.028 &$180eV$&$253eV$&$1.40$\\
\hline
1     &$-1.04$&$0.13$ &$0.10$&-0.046 &$162eV$&$234eV$&$1.44$\\
\hline
2     &$-0.97$&$0.05$ &$0.09$&-0.082 &$148eV$&$214eV$&$1.45$\\
\hline
3     &$-0.78$&$-0.06$&$0.08$&-0.107 &$128eV$&$189eV$&$1.48$\\
\hline
\end{tabular}


\begin{quote}
{Table 4: Values of the  physical
quantities for different subtraction points
(three times subtracted dispersion relations).}
\end{quote}
\end{center}

\mbox{}

\begin{center}
\begin{tabular}{|l|l|l|l|}
\hline
SP    &   $s_1$    &     $s_2$  \\
\hline
-10   &   $-9.8$   &    $-10.0$  \\
\hline
-6  &     $-5.6$   &  $-6.0$ \\
\hline
-4   &    $-3.7$     &  $-3.9$ \\
\hline
-2  &     $-1.9$     &  $-2.1$ \\
\hline
0   &     $1.3*10^{-3}$     &  $5*10^{-3}$ \\
\hline
1   &    $1.2$     &  $1.33$ \\
\hline
2   &    $1.9$     &  $2.1$ \\
\hline
3  &     $2.7$     &  $2.8$ \\
\hline
\end{tabular}


\begin{quote}
{Table 5: Values of the subtraction points $SP$ in units of
 $m_{\pi}^2$ for the twice subtracted dispersion
relation}
\end{quote}
\end{center}

\mbox{}

\begin{center}
\begin{tabular}{|l|l|l|l|l|l|l|}
\hline
SP    &$a$&$b$&$c$ & ${\it\Gamma}_{+-0}$ &${\it\Gamma}_{000}$ & $r$ \\
\hline
-10   &$-0.81$&$0.12$ &$0.04$&$410eV$&$624eV$&$1.52$\\
\hline
-6    &$-0.72$&$0.10$ &$0.06$&$277eV$&$423eV$&$1.53$\\
 \hline
-4    &$-0.77$&$0.14$ &$0.05$&$222eV$&$335eV$&$1.51$\\
\hline
-2    &$-0.90$&$0.20$ &$0.06$&$182eV$&$268eV$&$1.47$\\
\hline
0     &$-1.14$&$0.37$ &$0.03$&$145eV$&$203eV$&$1.40$\\
\hline
1     &$-0.96$&$0.28$ &$0.04$&$120eV$&$173eV$&$1.44$\\
\hline
2     &$-0.78$&$0.20$ &$0.05$&$95eV$&$140eV$&$1.47$\\
\hline
3     &$-0.71$&$-0.13$&$0.02$&$78eV$&$117eV$&$1.50$\\
\hline
\end{tabular}

\begin{quote}
{Table 6: Values of various physical observables for
different subtraction points and two subtractions.}
\end{quote}
\end{center}

\mbox{}

\begin{center}
\begin{tabular}{|l|l|l|}
\hline
                       & $\delta \bar c/\bar c=+ 25 \%$ & $\delta \bar c/\bar
c=- 25 \%$ \\
\hline
$\delta \bar d/\bar d=+ 25 \%$   & $0.15 + i 0.04$     & $-0.13 - i 0.03$ \\
\hline
$\delta \bar d/\bar d=- 25 \%$   & $0.18 + i 0.05$     & $-0.09 - i 0.02$ \\
\hline
\end{tabular}

\begin{quote}
{Table 7: Change of the reduced decay amplitude $\bar A$ at the center
of the physical decay region for a relative error of 25 \% on
constants $\bar c$, $\bar d$.}
\end{quote}\end{center}

\mbox{}

\mbox{}

\mbox{}

\noindent
{\LARGE\bf Figure captions}

\vspace{0.5 cm}

\begin{description}
\item[Fig. 1a]Integration contour for case i)
\item[Fig. 1b]Integration contour for case ii)
\item[Fig. 1c]Integration contour for case iii)
\item[Fig. 1d]Integration contour for case iv)
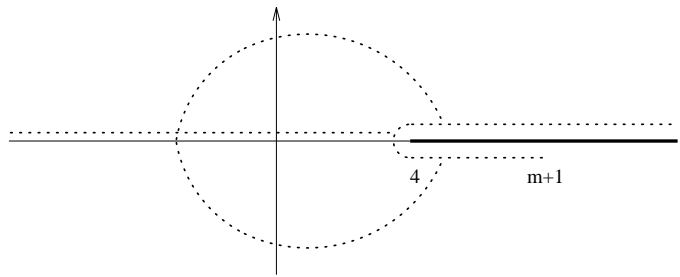
\item[Fig. 2 ]Lattice of points for the numerical
determination of the different functions
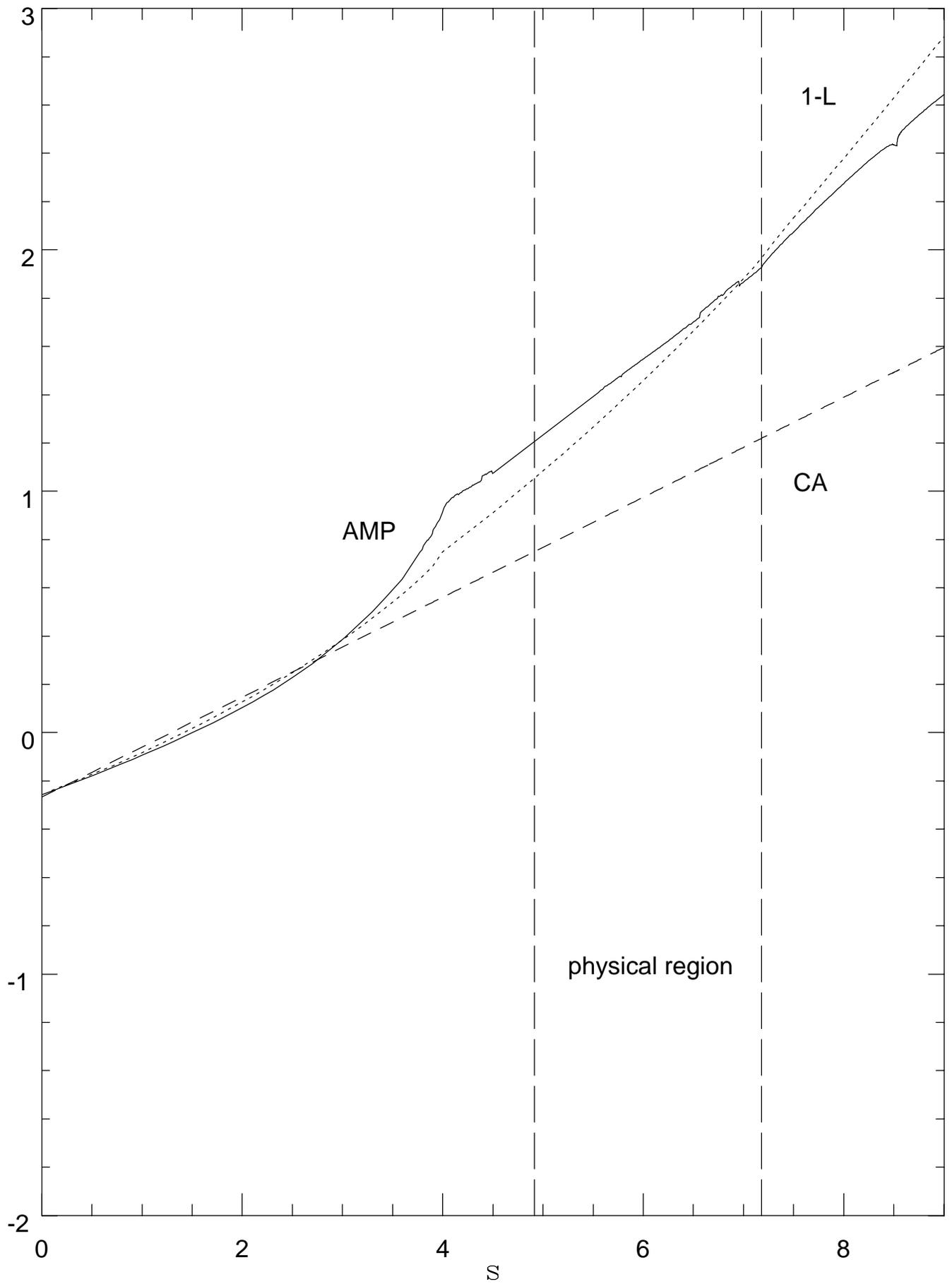
\item[Fig. 3 ] Real part of the decay amplitude for $s=u$ (in units of
$m_\pi^2$) for $SP=0$. The full line is the numerical solution of the coupled
integral
equations as described in the text. Also shown is the current algebra
prediction (dashed) and the chiral perturbation theory one-loop result
(dotted). The physical region lies between the two vertical lines.
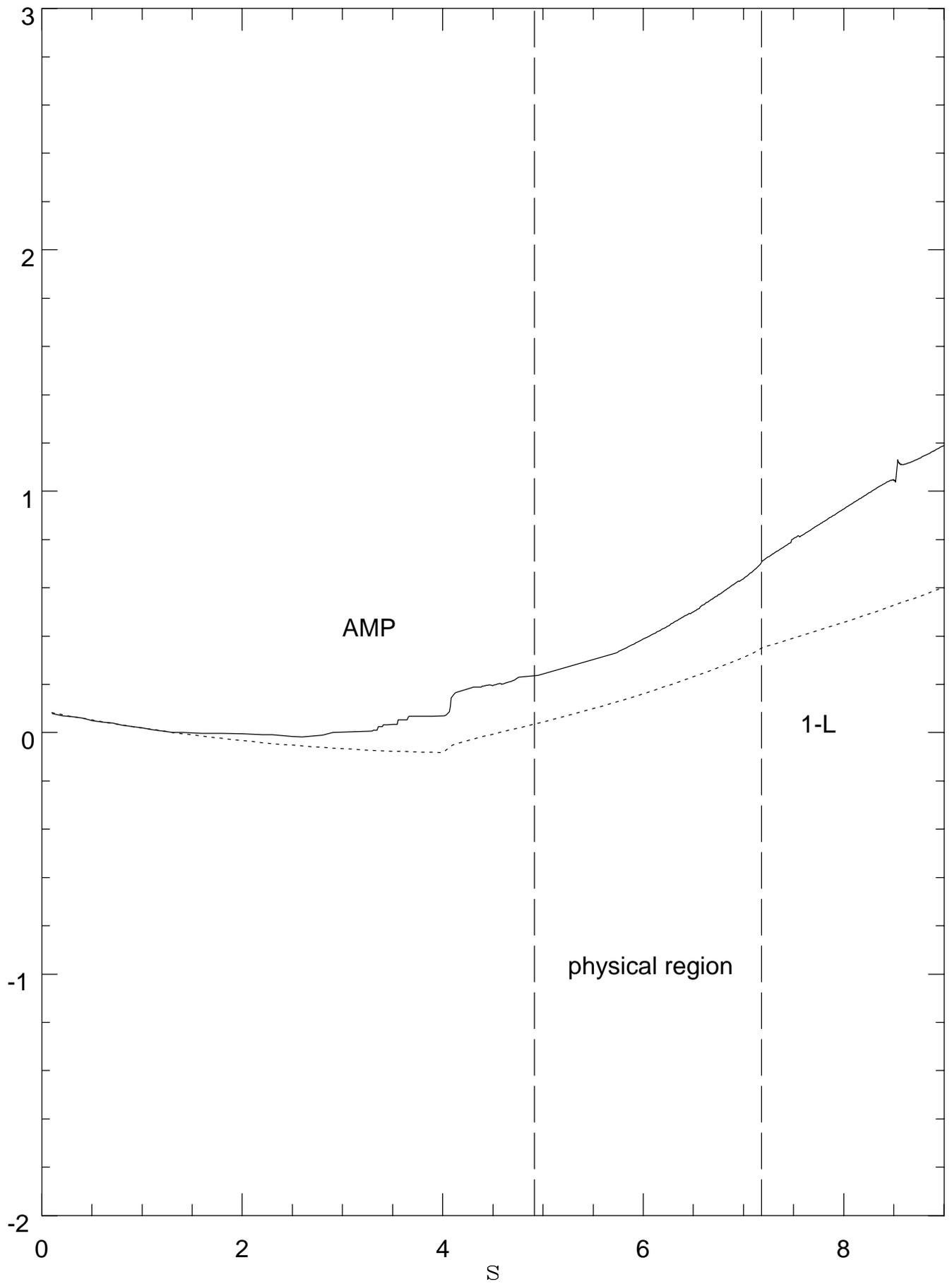
\item[Fig. 4 ] Imaginary part of the decay amplitude for $s=u$ (in units
of $m_\pi^2$) for $SP=0$. The full line is the result of the
numerical iteration
and the dotted one the chiral perturbation theory one-loop approximation.
\end{description}

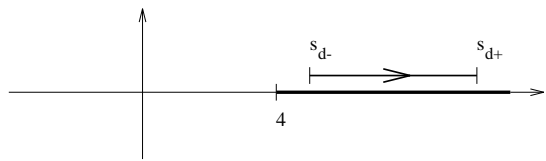
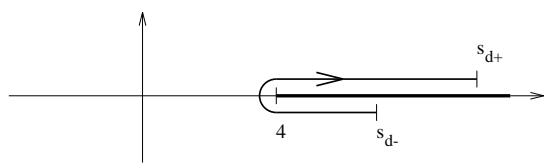
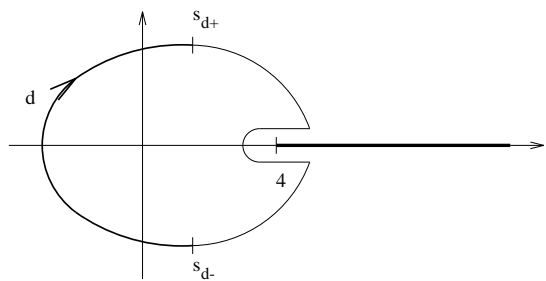
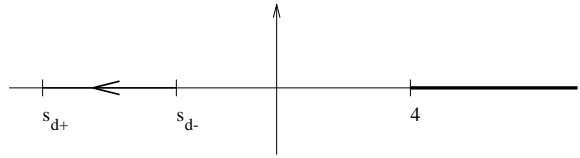
\end{document}